\documentclass[11pt,leqno]{article}

\usepackage{mathptmx}

\usepackage{amssymb,amsmath,amsthm}
\usepackage{comment}
\usepackage[latin1]{inputenc}    
\usepackage{pifont}    

\makeatletter%

\usepackage{nicefrac}
\usepackage{mathptmx}
\newcommand{\doublespacing}{\let\CS=
\@currsize\renewcommand{\baselinestretch}{1.75}\tiny\CS}
\newcommand{\extradoublespacing}{\let\CS=
\@currsize\renewcommand{\baselinestretch}{1.9}\tiny\CS}
\newcommand{\draftspacing}{\let\CS=
\@currsize\renewcommand{\baselinestretch}{2.0}\tiny\CS}
\newcommand{\hugedraftspacing}{\let\CS=
\@currsize\renewcommand{\baselinestretch}{2.4}\tiny\CS}
\makeatother%

\newcommand{\OMIT}[1]{} %
\newcommand{\gfootnote}[1]{} %

\newcommand\qedblob{\ding{113}}
\def\literalqed{{\ \nolinebreak\hfill\mbox{\qedblob\quad}}}

\newenvironment{proofs}{\noindent{\bf Proof.}\hspace*{1em}}{\literalqed\bigskip}

\newcommand{\seq}{\subseteq}

\newcommand{\scoreof}[1]{\mathit{score}(#1)}
\newcommand{\scoresub}[2]{\mathit{score}_{#1}(#2)}

\newcommand{\scoresublevel}[3]{\mathit{score}_{#1}^{#2}(#3)}

\newcommand{\tabscoresub}[1]{\mathit{score}_{#1}}

\newcommand{\hittingset}{\mbox{\rm Hitting Set}}

\newcommand{\p}{\mbox{\rm P}}

\newcommand{\np}{\mbox{\rm NP}}

\newcommand{\condition}{\,|\:}

\newenvironment{desctight}
  {\begin{list}{}{\setlength\labelwidth{0pt}%
        \setlength{\itemsep}{0.5pt}%
        \setlength{\parsep}{0pt}%
        \setlength\itemindent{-\leftmargin}%
        }}
    {\end{list}}

  \newtheorem{theorem}{Theorem}[section]
  
  \newtheorem{claim}[theorem]{Claim}

  \newtheorem{lemma}[theorem]{Lemma}

  \newtheorem{proposition}[theorem]{Proposition}
  \newtheorem{definition}[theorem]{Definition}
  
  \newtheorem{construction}[theorem]{Construction}
\usepackage{ulem}
\begin{document}

\title{Control Complexity in Fallback
Voting\thanks{A preliminary version of this paper appears in the proceedings
of \emph{Computing: the 16th Australasian Theory Symposium} (CATS-2010)
\cite{erd-rot:c:fallback-voting}.
Supported in part by the DFG under grants
RO~\mbox{1202/12-1} (within the European Science
Foundation's EUROCORES program LogICCC:
``Computational Foundations of Social Choice'') and RO~\mbox{1202/11-1}.
  Work done in part while the first author
  was visiting Universit\"{a}t Trier, while the third author
  was visiting the University of Rochester, and while the first and third
  author were visiting NICTA, Sydney, and the University of Newcastle.}}

\author{G\'{a}bor Erd\'{e}lyi\thanks{URL: 
\mbox{\tt{}ccc.cs.uni-duesseldorf.de/\mbox{\tiny$\sim\,$}erdelyi.}
}, \ 
Lena Piras,
\ and \ 
J\"{o}rg Rothe\thanks{URL: 
\mbox{\tt{}ccc.cs.uni-duesseldorf.de/\mbox{\tiny$\sim\,$}rothe.}
} \\
Institut f\"{u}r Informatik \\
Heinrich-Heine-Universit\"{a}t D\"{u}sseldorf \\
40225 D\"{u}sseldorf \\
Germany
}

\date{April 20, 2010}

\maketitle

\begin{abstract}
  We study the control complexity of fallback voting.  Like
  manipulation and bribery, electoral control describes ways of
  changing the outcome of an election; unlike manipulation or bribery
  attempts, control actions---such as adding/deleting/partitioning
  either candidates or voters---modify the participative structure of
  an election.  Via such actions one can try to either
  make a favorite candidate win (``constructive control'') or prevent
  a despised candidate from winning (``destructive control'').
  Computational complexity can be used to protect elections from
  control attempts, i.e., proving an election system resistant to some
  type of control shows that the success of the corresponding control
  action, though not impossible, is computationally prohibitive.

  We show that fallback voting, an election system combining approval
  with majority voting \cite{bra-san:j:preference-approval-voting}, is
  resistant to each of the
  common types of candidate control and
  to each
  common type of constructive control.  Among
  natural election systems with a polynomial-time winner
  problem, only plurality and sincere-strategy preference-based
  approval voting (SP-AV) were previously known to be fully resistant
  to candidate control
  \cite{bar-tov-tri:j:control,hem-hem-rot:j:destructive-control,erd-now-rot:j:sp-av},
  and only Copeland voting and SP-AV were previously known to be fully
  resistant to constructive control
  \cite{fal-hem-hem-rot:j:llull-copeland-full-techreport,erd-now-rot:j:sp-av}.
  However, plurality has fewer resistances to voter control,
  Copeland voting has fewer resistances to
  destructive control,
  and SP-AV
  (which like fallback voting has 19 out of 22 proven control resistances)
  is arguably less natural a system than fallback voting.
\end{abstract}
\maketitle                   %

\section{Introduction}
 \label{sec:introduction}

Voting is a way of aggregating individual preferences (or votes) to
achieve a societal consensus on which among several alternatives (or
candidates) to choose.  This is a central method of decision-making
not only in human societies but also in, e.g., multiagent systems
where autonomous software agents may have differing individual
preferences on a given number of alternatives.  Voting has been
studied intensely in areas as diverse as social choice theory and
political science, economics, operations research, artificial
intelligence, and other fields of computer science.  Voting
applications in computer science include the web-page ranking problem
\cite{dwo-kum-nao-siv:c:rank-aggregation}, similarity
search~\cite{fag-kum-siv:c:similarity-search},
planning~\cite{eph-ros:c:multiagent-planning}, and recommender
systems~\cite{gho-mun-her-sen:c:voting-for-movies}.  For such
applications, it is important to understand the computational
properties of election systems.

Various ways of tampering with the outcome of elections have been
studied from a complexity-theoretic perspective, in particular the
complexity of changing an election's outcome by \emph{manipulation}
\cite{bar-tov-tri:j:manipulating,bar-orl:j:polsci:strategic-voting,con-san-lan:j:when-hard-to-manipulate,hem-hem:j:dichotomy-scoring,fal-hem-hem-rot:c:single-peaked-preferences},
\emph{control}
\cite{bar-tov-tri:j:control,hem-hem-rot:j:destructive-control,fal-hem-hem-rot:j:llull-copeland-full-techreport,hem-hem-rot:j:hybrid,erd-now-rot:j:sp-av,fal-hem-hem-rot:c:single-peaked-preferences},
and \emph{bribery}
\cite{fal-hem-hem:j:bribery,fal-hem-hem-rot:j:llull-copeland-full-techreport},
see also the surveys by Faliszewski et al.~\cite{fal-hem-hem-rot:b:richer}
and Baumeister
et al.~\cite{bau-erd-hem-hem-rot:t:computational-apects-of-approval-voting}.

In control scenarios, an external actor---commonly referred to as the
``chair''---seeks to either make a favorite candidate win
(constructive control) or block a despised candidate's victory
(destructive control) via actions that change the participative
structure of the election.  Such actions include adding, deleting, and
partitioning either candidates or voters; the $22$ commonly studied
control actions, and their corresponding control problems, are
described formally in Section~\ref{sec:preliminaries}.

We study the control complexity of fallback voting, an election system
introduced by Brams and Sanver~\cite{bra-san:j:preference-approval-voting}
as a way of combining approval and preference-based voting.  We prove
that fallback voting is resistant (i.e., the corresponding control
problem is $\np$-hard) to each of the $14$ common types of candidate
control.  In addition, we show that fallback voting is resistant to
five types of voter control.  In particular, fallback voting is resistant
to each of the $11$ common types of constructive control.  Among natural
election systems with a polynomial-time winner determination procedure,
only plurality and sincere-strategy preference-based approval voting
(SP-AV) were previously known to be fully resistant to candidate control
\cite{bar-tov-tri:j:control,hem-hem-rot:j:destructive-control,erd-now-rot:j:sp-av}, and only Copeland voting and
SP-AV were previously known to be fully resistant to constructive control
\cite{fal-hem-hem-rot:j:llull-copeland-full-techreport,erd-now-rot:j:sp-av}.
However, SP-AV (as
modified by \cite{erd-now-rot:j:sp-av}) is arguably less natural a
system than fallback voting,\footnote{SP-AV is (a variant of) 
another hybrid system combining approval and
preference-based voting that was also proposed by Brams and Sanver
(see \cite{bra-san:j:critical-strategies-under-approval}).  The reason we
said SP-AV is less natural than fallback voting is that, in order to
preserve
``admissibility'' of votes (as required by Brams and Sanver
\cite{bra-san:j:critical-strategies-under-approval} to preclude
trivial approval strategies), SP-AV (as modified by Erd\'{e}lyi et
al.~\cite{erd-now-rot:j:sp-av}) employs an additional rule to (re-)coerce
admissibility (in particular, if in the course of a control action an
originally admissible vote becomes inadmissible).  This point has been
discussed in detail by Baumeister et 
al.~\cite{bau-erd-hem-hem-rot:t:computational-apects-of-approval-voting}.
In a nutshell, this rule, if
applied, changes the approval strategies of the votes originally cast
by the voters.  The effect of this rule is that SP-AV can be seen as a
hybrid between approval and plurality voting, and it indeed possesses
each resistance either of these two systems has (and many of these
resistance proofs are based on slightly modified constructions from
the resistance proofs for either plurality or approval due to
Hemaspaandra et al.~\cite{hem-hem-rot:j:destructive-control}).  In contrast,
here we study the original variant of fallback voting, as proposed by
Brams and Sanver~\cite{bra-san:j:preference-approval-voting}, in which
votes, once
cast, do not change.
}
and plurality has fewer resistances to
voter control and Copeland voting has fewer resistances to
destructive control than fallback voting.

This paper is organized as follows.  In
Section~\ref{sec:preliminaries}, we recall some notions from voting
theory, define the commonly studied types of control, and explain
Brams and Sanver's fallback voting
procedure~\cite{bra-san:j:critical-strategies-under-approval} in
detail.  Our results on the control complexity of fallback voting are
presented in Section~\ref{sec:results}.
Finally, Section~\ref{sec:conclusions} provides some conclusions and
open questions.

\section{Preliminaries}
\label{sec:preliminaries}

\subsection{Elections and Electoral Control}

An election is a pair $(C,V)$, where $C$ is a finite set of candidates
and $V$ is a collection of votes over~$C$.  How the votes are
represented depends on the election system used.  Many systems (such
as majority, Condorcet, and Copeland voting as well as the class
of scoring protocols including plurality, veto, and Borda
voting; see, e.g., \cite{bra-fis:b:voting-procedures})
represent the voters' preferences as strict, linear orders
over the candidates.  In approval voting
\cite{bra-fis:j:approval-voting,bra-fis:b:approval-voting,bra:j:approval-multicandidate},
however, a vote over $C$ is a so-called approval
strategy, a subset of $C$ containing all candidates approved of by
this voter, whereas he or she disapproves of all other candidates.  As
is standard, we assume that collections $V$ of votes over $C$ are
given as lists of ballots, i.e., if, say, five voters
express the same preference (be it as a linear order or as an
approval strategy or both) then this vote occurs five times in~$V$.

An election system is a rule determining the winner(s) of a given
election $(C,V)$.  For example, in \emph{plurality-rule
  elections}, the winners are precisely those candidates in $C$ who
are ranked first place by the most voters.  In \emph{(simple)
  majority-rule elections}, the winner is that candidate in $C$ (assuming
one exists) who is ranked first place by a strict majority of the
votes (i.e., by more than $\nicefrac{\|V\|}{2}$ voters).  In
\emph{approval voting}, every candidate scores one point for each
approval by a voter, and whoever has the most points wins.

All our control problems come in two variants: The \emph{constructive}
case \cite{bar-tov-tri:j:control} focuses on the chair seeking to make
a favorite candidate win; the \emph{destructive} case
\cite{hem-hem-rot:j:destructive-control} is concerned with the chair
seeking to make a despised candidate not win.  Due to space, we
describe these control problems only very briefly.  Detailed, formal
problem descriptions, along with many motivating examples and
scenarios where these types of control naturally occur in real-world
elections, can be found, e.g., in
\cite{bar-tov-tri:j:control,hem-hem-rot:j:destructive-control,fal-hem-hem-rot:j:llull-copeland-full-techreport,hem-hem-rot:j:hybrid,erd-now-rot:j:sp-av,bau-erd-hem-hem-rot:t:computational-apects-of-approval-voting}.
As an example, we state one such problem formally:
\begin{desctight}

\item[Name:] Constructive Control by Adding a Limited Number of Candidates.

\item[Given:] An election $(C \cup D, V)$, $C \cap D = \emptyset$, a
  distinguished candidate $c \in C$, and a nonnegative integer~$k$.
  ($C$ is the set of originally qualified candidates and $D$ is the
  set of spoiler candidates that may be added.)

\item[Question:] Does there exist a subset $D' \seq D$ such that $\| D' \|
  \leq k$ and $c$ is the unique winner (under the election system at
  hand) of election $(C \cup D', V)$?

\end{desctight}

Constructive Control by Adding an Unlimited Number of Candidates is
the same except there is no limit $k$ on the number of spoiler
candidates that may be added.  The destructive variants of both
problems are obtained by asking whether $c$ is \emph{not} a unique
winner of $(C \cup D', V)$.  The constructive control by deleting
candidates problem is defined analogously except that we are given an
election $(C,V)$, a candidate $c \in C$, and a deletion limit~$k$, and
ask whether $c$ can be made a unique winner by deleting up to $k$
candidates from~$C$.  The destructive version of this problem is the
same except that we now want to preclude $c$ from being a unique winner
(and we are not allowed to delete~$c$).

Constructive Control by Partition (or Run-Off Partition) of Candidates
takes as input an election $(C,V)$ and a candidate $c \in C$ and asks
whether $c$ can be made a unique winner in a certain two-stage
election consisting of one (or two) first-round subelection(s) and a
final round.  In both variants, following
\cite{hem-hem-rot:j:destructive-control}, we consider two
tie-handling rules, TP (``ties promote'') and TE (``ties eliminate''),
that enter into force when more candidates than one are tied for
winner in any of the first-round subelections.  In the variant with
run-off and under the TP rule, the question is whether $C$ can be
partitioned into $C_1$ and $C_2$ such that $c$ is the unique winner of
election $(W_1 \cup W_2,V)$, where~$W_i$, $i \in \{1,2\}$, is the set
of winners of subelection $(C_i,V)$.  In the variant without run-off
(again under TP), the question is whether $C$ can be partitioned into
$C_1$ and $C_2$ such that $c$ is the unique winner of election
$(W_1 \cup C_2,V)$.  In both cases, when the TE rule is used, none of
multiple, tied first-round subelection winners is promoted to the
final round (e.g., if we have a run-off and $\|W_2\| \geq 2$ then the
final-round election collapses to $(W_1,V)$; only a unique first-round
subelection winner is promoted to the final round).  It is obvious how to
obtain the destructive variants of these eight problems formalizing
control by candidate partition.  Summing up, we now have defined $14$
candidate control problems.

Constructive Control by Adding Voters is the problem of deciding,
given an election $(C,V \cup V')$, $V \cap V' = \emptyset$, where $V$ is
a collection of registered voters and $V'$ a pool of as yet
unregistered voters that can be added, a candidate $c \in C$, and an
addition limit~$k$, whether there is a subset $V'' \subseteq V'$ of size
at most $k$ such that $c$ is the unique winner of election $(C,V \cup V'')$.
Constructive Control by Deleting Voters asks, given an election
$(C,V)$, a candidate $c \in C$, and a deletion limit~$k$, whether it
is possible to make $c$ the unique winner by deleting up to $k$ votes
from~$V$.  For the TP
rule, in Constructive Control by Partition of Voters the question is,
given an election $(C,V)$ and a candidate $c \in C$, whether $V$ can
be partitioned into $V_1$ and $V_2$ such that $c$ is the unique winner
of election $(W_1 \cup W_2,V)$, where~$W_i$, $i \in \{1,2\}$, is the
set of winners of subelection $(C,V_i)$.  It is obvious how this
problem definition changes when the TE
rule is used, and also how the destructive variants of these four
voter control problems are obtained.  Summing up, we now have defined
eight voter control problems and thus a total of $22$ control
problems.
These problems are due to \cite{bar-tov-tri:j:control} and
\cite{hem-hem-rot:j:destructive-control} (see also
\cite{fal-hem-hem-rot:j:llull-copeland-full-techreport}), and their
complexity has been studied for a variety of election systems, with
the goal of using complexity as a barrier that makes attempts of
changing the outcome of an election via control, although not
impossible, at least computationally prohibitive.

We assume the reader is familiar with the standard complexity classes
$\p$ (deterministic polynomial time) and $\np$ (nondeterministic
polynomial time) and with the notions of hardness and completeness. In
particular, a problem $B$ is said to be $\np$-hard if for each $A \in
\np$ it holds that $A$ (polynomial-time many-one) reduces to~$B$,
where we say $A$ reduces to $B$ if there is a polynomial-time function
$\varphi$ such that for each input $x$, $x$ is in $A$ if and only if
$\varphi(x)$ is in~$B$.  Every $\np$-hard problem in $\np$ is said to
be $\np$-complete.  For more background on complexity theory, we refer
to the
textbooks~\cite{gar-joh:b:int,pap:b-1994:complexity,rot:b:cryptocomplexity}.

Given a control type $\Phi$, some election systems have the
advantageous property that it is never possible for the chair to reach
his or her goal of exerting control of type~$\Phi$.  In such a case,
the system is said to be \emph{immune to~$\Phi$}; otherwise, the
system is said to be \emph{susceptible to~$\Phi$}.  We say an election
system $\mathcal{E}$ is \emph{resistant to~$\Phi$} if $\mathcal{E}$ is
susceptible to $\Phi$ and the control problem corresponding to $\Phi$
is $\np$-hard.  If, however, $\mathcal{E}$ is susceptible to $\Phi$
and the control problem corresponding to $\Phi$ is in $\p$, we say
$\mathcal{E}$ is \emph{vulnerable to~$\Phi$}.

\subsection{Fallback Voting}

Brams and Sanver 
proposed two hybrid systems that combine approval voting and systems
based on linear preferences, \emph{sincere-strategy preference-based
  approval voting}~\cite{bra-san:j:critical-strategies-under-approval}
and \emph{fallback voting} (FV)~\cite{bra-san:j:preference-approval-voting}.
Erd\'{e}lyi et al.~\cite{erd-now-rot:j:sp-av} studied the
control complexity of SP-AV, as they dubbed their variant of the election
system originally proposed by
\cite{bra-san:j:critical-strategies-under-approval}.  As explained in
the last paragraph of the introduction, SP-AV combines, in a certain
sense, approval with plurality voting.  In this paper, we
investigate the election system FV
with respect to
electoral control.  FV can be thought of as combining, in a certain
sense, approval with majority voting.  Unlike in SP-AV, in FV only the
approved candidates are ranked by a tie-free linear order.

\begin{definition}[\cite{bra-san:j:preference-approval-voting}]
\label{def:fv}
Let $(C,V)$ be an election.
Every voter $v \in V$ provides a subset $S_v \subseteq C$ indicating
that $v$ approves of all candidates in $S_v$ and disapproves of all
candidates in $C-S_v$.  $S_v$ is called \emph{$v$'s approval
  strategy}.  In addition, each voter $v\in V$ provides a tie-free
linear ordering of all candidates in~$S_v$.

If $S_v = \{c_1, c_2,\ldots , c_k\}$ and $v$ ranks the candidates in
$S_v$ by $c_1 > c_2 > \cdots > c_k$, where $c_1$ is $v$'s most
preferred candidate, $c_2$ is $v$'s second most preferred candidate,
etc., and $c_k$ is $v$'s least preferred candidate (among the candidates
$v$ approves of), we denote the vote $v$ by
\[
\begin{array}{c@{\ \ }c@{\ \ }c@{\ \ }c@{\ \ }c@{\ \ }c}
 c_1 & c_2 & \cdots & c_k & | & C - S_v,
\end{array}
\]
where the approved candidates to the left of the approval line are
ranked and the disapproved candidates in $C - S_v$ are written as an
(unordered) set to the right of the approval line.
For each $c \in C$, let
\[
\scoresub{(C,V)}{c} = \|\{v \in V \condition c \in S_v\}\|
\]
denote the number of $c$'s approvals in $(C,V)$, and let
$\scoresublevel{(C,V)}{i}{c}$ be the \emph{level~$i$ score of $c$ in
  $(C,V)$}, which is the number of $c$'s approvals when ranked between
the (inclusively) first and $i^{th}$ position.
Winner determination in FV is based on determining level~$i$ FV
winners as follows:
\begin{enumerate} 
 \item On the first level, only the highest ranked approved candidates
   are considered in each voters' approval strategy. If a candidate
   $c\in C$ has a strict majority on this level (i.e.,
   $\scoresublevel{(C,V)}{1}{c} > \nicefrac{\|V\|}{2}$), then $c$ is
   the \emph{(unique) level~$1$ FV winner of the election}.
 
\item On the second level, if there is no level~$1$ winner, the two
  highest ranked approved candidates (if they exist) are considered in
  each voters' approval strategy. If there is exactly one candidate
  $c\in C$ who has a strict majority on this level (i.e.,
  $\scoresublevel{(C,V)}{2}{c} > \nicefrac{\|V\|}{2}$), then $c$ is
  the \emph{(unique) level~$2$ FV winner of the election}.  If there
  are at least two such candidates, then every candidate with the
  highest level~$2$ score is a \emph{level~$2$ FV winner of the
    election}.
 
\item If there is no level~$1$ or level~$2$ FV winner, we in this way
  continue level by level until there is at least one candidate who
  was approved by a strict majority on the current level, say
  level~$i$.  If there is only one such candidate, he or she is the
  \emph{(unique) level~$i$ FV winner of the election}.  If there are
  at least two such candidates, then every candidate with the highest
  level~$i$ score is a \emph{level~$i$ FV winner of the election}.
 
 \item For each $c \in C$, if $c$ is a level~$i$ FV winner of $(C,V)$
   for some (smallest) $i \leq \|C\|$ then $c$ is said to be an
   \emph{FV winner of $(C,V)$}.
   Otherwise (i.e., if for no $i \leq \|C\|$ there is
   a level~$i$ FV winner), every candidate with the highest
   $\scoresub{(C,V)}{c}$ is an \emph{FV winner of~$(C,V)$}.
\end{enumerate}
\end{definition}

Note that if there exists a level~1 FV winner, he or she is always the
election's unique FV winner.  In contrast to SP-AV (where
\emph{admissibility} of the votes---$\emptyset \neq S_v \neq C$ for
each $v \in V$---is coerced by moving the approval line whenever
necessary, see~\cite{erd-now-rot:j:sp-av}), in fallback voting no
changes are made to the ballots, regardless of the control action
taken.  That is, we don't require votes to be admissible, i.e., both
empty ($S_v=\emptyset$) and full ($S_v=C$) approval strategies are
allowed.  By definition, votes in an FV election are always
\emph{sincere} (i.e., if a voter $v$ approves of a candidate $c$ then
the voter also approves of all candidates ranked higher than~$c$).  In
contrast, sincerity has to be enforced in SP-AV by declaring insincere
votes void.

\begin{table*}[t!]
\centering
{\small
\begin{tabular}{|l||l|l||l|l||l|l|}
\hline
                    & \multicolumn{2}{c||}{FV}
                    & \multicolumn{2}{c||}{SP-AV}
                    & \multicolumn{2}{c|}{AV}
\\ \cline{2-7}
Control by          & Const. & Dest.
                    & Const. & Dest.
                    & Const. & Dest.
\\ \hline\hline
Adding an Unlimited Number of Candidates
                    & {\bf R}               & {\bf R} 
                    & 	   R                &      R 
                    &      I                &      V 
\\ \hline
Adding a Limited Number of Candidates
                    & {\bf R}               & {\bf R} 
                    &      R                &      R 
                    &      I                &      V 
\\ \hline
Deleting Candidates & {\bf R}               & {\bf R} 
                    &      R                &      R 
                    &      V                &      I 
\\ \hline
Partition of Candidates
                    & {\bf TE: R}           & {\bf TE: R} 
                    &      TE: R            &      TE: R 
                    &      TE: V            &      TE: I 
\\
                    & {\bf TP: R}           & {\bf TP: R} 
                    &      TP: R            &      TP: R      
                    &      TP: I            &      TP: I 
\\ \hline
Run-off Partition of Candidates
                    & {\bf TE: R}           & {\bf TE: R} 
                    &      TE: R            &      TE: R 
                    &      TE: V            &      TE: I 
\\ 
                    & {\bf TP: R}           & {\bf TP: R} 
                    &      TP: R            &      TP: R      
                    &      TP: I            &      TP: I 
\\ \hline
Adding Voters       & {\bf R}               & {\bf V}         
                    &      R                &      V 
                    &      R                &      V 
\\ \hline
Deleting Voters     & {\bf R}               & {\bf V}      
                    &      R                &      V 
                    &      R                &      V 
\\ \hline
Partition of Voters
                    &      {\bf TE: R}      &      {\bf TE: S} 
                    &      TE: R            &      TE: V 
                    &      TE: R            &      TE: V 
\\
                    &      {\bf TP: R}      &      {\bf TP: R}   
                    &      TP: R            &      TP: R      
                    &      TP: R            &      TP: V 
\\ 
\hline
\end{tabular}
}
\caption{\label{tab:summary-of-results}
Overview of results.  Key: I means immune,
S means susceptible,
R means resistant,
V means vulnerable,
TE means ties eliminate, and TP means ties promote.
Results new to this paper are in boldface.  Results
for approval voting are due to \cite{hem-hem-rot:j:destructive-control}
and results for SP-AV are due to \cite{erd-now-rot:j:sp-av}.
} 
\end{table*}

\section{Results}
\label{sec:results}

\subsection{Overview}
\label{sec:results:overview}

Theorem~\ref{thm:FV-summary-of-results} and
Table~\ref{tab:summary-of-results} show the results on the control
complexity of fallback voting.

\begin{theorem}
\label{thm:FV-summary-of-results}
Fallback voting is resistant, vulnerable, and susceptible to the $22$
types of control defined in Section~\ref{sec:preliminaries} as shown
in Table~\ref{tab:summary-of-results}.
\end{theorem}

\subsection{Susceptibility}
\label{sec:results:susceptibility}

Among the $22$ control types we consider, approval voting has nine
immunities \cite{hem-hem-rot:j:destructive-control}, see
Table~\ref{tab:summary-of-results}.  Some of these immunities
immediately follow from the unique version of the Weak Axiom of
Revealed Preference (Unique-WARP, for short), which says that if a
candidate $c$ is the unique winner of an election $(C,V)$ then $c$ is
also the unique winner of each election $(C',V)$
such that $C' \subset C$ and $c \in C'$ (where, for convenience, we
use the same symbol $V$ but view the preferences in $V$ as being
restricted to the candidates in~$C'$; this convention is also adopted
when we speak of subelections in the context of candidate-control problems).

Unlike approval voting but just as SP-AV, fallback voting does not
satisfy Unique-WARP.

\begin{proposition}
\label{prop:unique-warp}
Fallback voting does not satisfy Unique-WARP.
\end{proposition}

\begin{proofs}
Consider the election $(C,V)$ with candidate set $C = \{a, b, c,
d\}$ and voter collection $V = \{v_1, v_2, \ldots , v_6\}$:
\[
\begin{array}{rc@{\ \ }c@{\ \ }c@{\ \ }c@{\ \ }c}
 & \multicolumn{5}{c}{(C,V)} \\ 
\cline{2-6}
v_1=v_2=v_3: & a & c &   & | & \{b,d\}   \\
v_4=v_5:     & b & d & c & | & \{ a \}   \\
v_6:         & d & a & c & | & \{ b \}   \\
\end{array}
\]

There is no level~1 FV winner, and the unique level~2 FV winner of the
election $(C,V)$ is candidate $a$ with
$\scoresublevel{(C,V)}{2}{a}=4$.  By removing candidate $b$ from the
election, we get the subelection $(C',V)$ with $C'=\{a,c,d\}$.  (Recall
that, after removing~$b$, $V$ is viewed as restricted to~$C'$; e.g.,
voter $v_4$ in $V$ is now changed to
$\begin{array}{@{\, }c@{\ \ }c@{\ \ }c@{\ \ }c@{\, }} d & c
  & | & \{a\}
\end{array}$.)
There is again no level~1 FV winner in $(C',V)$.  However, there are
two candidates on the second level with a strict majority, namely
candidates $a$ and~$c$.  Since $\scoresublevel{(C',V)}{2}{c}=5$ is
greater than $\scoresublevel{(C',V)}{2}{a}=4$, the unique level~2 FV
winner of the subelection $(C',V)$ is candidate~$c$.  Thus, FV does
not satisfy Unique-WARP.~\end{proofs}

Indeed, as we will now show, fallback voting is susceptible to each of
our $22$ control types.  Our proofs make use of the results of
\cite{hem-hem-rot:j:destructive-control} that provide general proofs
of and links between certain susceptibility cases.  For the sake of
self-containment, we state their results, as
Theorems~\ref{thm:voiced-control},
\ref{thm:duality-constructive-destructive-control},
and~\ref{thm:susceptibility-implications}, in the appendix.

We start with susceptibility to candidate control.

\begin{lemma}
 \label{lem:susceptible-candidate-control}
Fallback voting is susceptible to constructive and destructive control
by adding candidates (in both the ``limited'' and the ``unlimited''
case), by deleting candidates, and by partition of candidates (with or
without run-off and for each in both model TE and model TP).
\end{lemma}

\begin{proofs}
From Theorem~\ref{thm:voiced-control} and the fact that FV is a voiced
voting system,\footnote{An election system is said to be \emph{voiced}
  if the single candidate in any one-candidate election always wins.}
it follows that FV is susceptible to constructive control by deleting
candidates, and to destructive control by adding candidates (in both
the ``limited'' and the ``unlimited'' case).

Now, consider the election $(C,V)$ given in the proof of
Proposition~\ref{prop:unique-warp}.  The unique FV winner of the
election is candidate $a$.  Partition $C$ into $C_1=\{a,c,d\}$ and
$C_2=\{b\}$. The unique FV winner of subelection $(C_1,V)$ is
candidate~$c$, as shown in the proof of
Proposition~\ref{prop:unique-warp}.  In both partition and run-off
partition of candidates and for each in both tie-handling models, TE
and TP, candidate $b$ runs against candidate $c$ in the final stage of
the election. The unique FV winner is in each case candidate
$c$. Thus, FV is susceptible to destructive control by partition of
candidates (with or without run-off and for each in both model TE and
model~TP).

By Theorem~\ref{thm:susceptibility-implications}, FV is also
susceptible to destructive control by deleting candidates. By
Theorem~\ref{thm:duality-constructive-destructive-control}, FV is also
susceptible to constructive control by adding candidates (in both the
``limited'' and the ``unlimited'' case).

Now, changing the roles of $a$ and $c$ makes $c$ our distinguished
candidate. In election $(C,V)$, $c$ loses against candidate $a$. By
partitioning the candidates as described above, $c$ becomes the unique
FV winner of the election. Thus, FV is susceptible to constructive
control by partition of candidates (with or without run-off and for
each in both tie-handling models, TE and TP).
\end{proofs}

We now turn to susceptibility to voter control.

\begin{lemma}
 \label{lem:susceptible-voter-control}
Fallback voting is susceptible to constructive and destructive 
control by adding voters, by deleting voters, and by partition 
of voters (in both model TE and model TP).
\end{lemma}

\begin{proofs}
Consider the election $(C,V)$, where $C=\{a,b,c,d\}$ is the 
set of candidates and $V=\{v_1,v_2,v_3,v_4\}$ is the 
collection of voters with the following preferences:
\[
\begin{array}{lc@{\ \ }c@{\ \ }c@{\ \ }c@{\ \ }c}
 & \multicolumn{5}{c}{(C,V)} \\ \cline{2-6}
v_1: & a & c &   & | & \{b,d\} \\
v_2: & d & c &   & | & \{a,b\} \\
v_3: & b & a & c & | & \{ d \} \\
v_4: & b & a &   & | & \{c,d\} \\
\end{array}
\]
We partition $V$ into $V_1=\{v_1,v_2\}$ and $V_2=\{v_3,v_4\}$. Thus we
split $(C,V)$ into two subelections:
\[
\begin{array}{lc@{\ \ }c@{\ \ }c@{\ \ }c@{\ \ }c@{\ \ }c@{\ \ }c@{\ \ }c@{\ \ }c@{\ \ }c@{\ \ }c}
 & \multicolumn{5}{c}{(C,V_1)} & \mbox{\quad and\quad }
 & \multicolumn{5}{c}{(C,V_2)}
\\ \cline{2-6}\cline{8-12}
v_1: & a & c &   & | & \{b,d\} &
     &   &   &   &   &         \\
v_2: & d & c &   & | & \{a,b\} & 
     &   &   &   &   &         \\
v_3: &   &   &   &   &         &
     & b & a & c & | & \{ d \} \\
v_4: &   &   &   &   &         &
     & b & a &   & | & \{c,d\} \\
\end{array}
\]

Clearly, candidate $a$ is the unique  level~2
FV winner of $(C,V)$.
However, $c$ is the unique level~2 FV winner of $(C,V_1)$ and $b$
is the unique level~1 FV winner of $(C,V_2)$, and so $a$ is not 
promoted to the final stage. Thus, FV is susceptible
to destructive control by partition of voters in
both tie-handling models, TE and TP.

By Theorem~\ref{thm:voiced-control} and the fact that FV is a voiced
voting system, FV is susceptible to destructive control by deleting
voters.  By
Theorem~\ref{thm:duality-constructive-destructive-control}, FV is also
susceptible to constructive control by adding voters.

By changing the roles of $a$ and $c$ again, we can see that FV is
susceptible to constructive control by partition of voters in both
model TE and model~{TP}. By
Theorem~\ref{thm:susceptibility-implications}, FV is also susceptible
to constructive control by deleting voters.  Finally, again by
Theorem~\ref{thm:duality-constructive-destructive-control}, FV is
susceptible to destructive control by adding voters.~\end{proofs}

\subsection{Candidate Control}
\label{sec:results:candidate-control}

All resistance results in this section follow via
Lemma~\ref{lem:susceptible-candidate-control}, showing susceptibility,
and a reduction from the well-known $\np$-complete problem
$\hittingset$~\cite{gar-joh:b:int}, showing $\np$-hardness of the
corresponding control problem.  $\hittingset$ is defined as follows:

\begin{desctight}

\item[Name:] $\hittingset$.

\item[Instance:] A set $B=\{b_1,b_2,\ldots , b_m\}$, a collection
  $\mathcal{S} =\{S_1,S_2,\ldots , S_n\}$ of nonempty subsets
  $S_i\subseteq B$, and a positive integer $k\leq m$.

\item[Question:] Does $\mathcal{S}$ have a hitting set of size at most
$k$, i.e., is there a set $B'\subseteq B$ with $\|B'\|\leq k$ such that
for each $i$, $S_i \cap  B'\neq \emptyset$?  
\end{desctight}

We now show that fallback voting is resistant to all types of
constructive and destructive candidate control defined in
Section~\ref{sec:preliminaries}.  To this end, we present a general
construction that will be applied (in
Theorems~\ref{thm:adding-candidates},
\ref{thm:deleting-candidates-destr},
and~\ref{thm:partition-candidates} below) to all these control
scenarios except constructive control by deleting candidates (which
will be handled via a different construction in
Theorem~\ref{thm:deleting-candidates-constr}).

\begin{construction}
\label{con:resistance-general-candidate-control}
Let $(B,\mathcal{S},k)$ be a given instance of $\hittingset$, where $B
= \{b_1, b_2, \ldots , b_m\}$ is a set, $\mathcal{S} = \{S_1, S_2,
\ldots , S_n\}$ is a collection of nonempty subsets $S_i \seq B$, and
$k$ is a positive integer.  Without loss of generality, we may assume
that $n>1$ (since $\hittingset$ is trivially solvable for $n \leq 1$)
and that $k < m$ (since $B$ is always a hitting set of size $k$ if $m
= k$ when $\mathcal{S}$ contains only nonempty sets).

Define the election $(C,V)$, where $C = B \cup \{c,d,w\}$ is the
candidate set and where $V$ consists of the following $6n(k+1)+4m+11$
voters:
\begin{enumerate}
\item There are $2m+1$ voters of the form:
\[
\begin{array}{c@{\ \ }c@{\ \ }c}
 c & | & B\cup \{d,w\}.
\end{array}
\]
\item There are $2n+2k(n-1)+3$ voters of the form:
\[
\begin{array}{c@{\ \ }c@{\ \ }c@{\ \ }c}
 c & w & | & B\cup \{d\}.
\end{array}
\]
\item There are $2n(k+1)+5$ voters of the form:
\[
\begin{array}{c@{\ \ }c@{\ \ }c@{\ \ }c}
 w & c & | & B\cup \{d\}.
\end{array}
\]
\item For each $i$, $1\leq i \leq n$, there are $2(k+1)$ voters of the
  form:\footnote{As a notation, when a vote contains a set of approved
    candidates, such as
$\begin{array}{@{\, }c@{\ \ }c@{\ \ }c@{\ \ }c@{\ \ }c@{\, }}
 c & D & a & | & C - (D \cup \{a,c\})
\end{array}$ for a subset $D \subseteq C$ of the candidates,
this is a shorthand for
$\begin{array}{@{\, }c@{\ \ }c@{\ \ }c@{\ \ }c@{\ \ }c@{\ \ }c@{\ \ }c@{\, }}
c & d_1 & \cdots & d_{\ell} & a & | & C - (D \cup \{a,c\})
\end{array}$,
where the elements of $D = \{d_1, \ldots , d_{\ell}\}$ are ranked with
respect to a (tacitly assumed) fixed ordering of all candidates in~$C$.}
\[
\begin{array}{c@{\ \ }c@{\ \ }c@{\ \ }c@{\ \ }c}
d & S_i & c & | & (B-S_i)\cup \{w\}.
\end{array}
\]
\item For each $j$, $1\leq j \leq m$, there 
are two voters of the form:
\[
\begin{array}{c@{\ \ }c@{\ \ }c@{\ \ }c@{\ \ }c}
d & b_j & w & | & (B-\{b_j\})\cup \{c\}.
\end{array}
\]
\item There are $2(k+1)$ voters of the form:
\[
\begin{array}{c@{\ \ }c@{\ \ }c@{\ \ }c@{\ \ }c}
 d & w & c & | & B.
\end{array}
\]
\end{enumerate}
\end{construction}

It is easy to see that there is no level~1 FV winner in election
$(\{ c,d,w \} ,V)$ and that we have the following level~2 scores
in this election:
\begin{eqnarray*}
\scoresublevel{(\{ c,d,w \} ,V)}{2}{c} & = & 2(m-k)+6n(k+1)+9,  \\
\scoresublevel{(\{ c,d,w \} ,V)}{2}{d} & = & 2n(k+1)+2(m+k+1),  \mbox{ and} \\
\scoresublevel{(\{ c,d,w \} ,V)}{2}{w} & = & 4n(k+1)+2m+10.
\end{eqnarray*}
Thus, $c$ is the unique level~2 FV winner of $(\{ c,d,w \} ,V)$.

The proofs of Theorems~\ref{thm:adding-candidates},
\ref{thm:deleting-candidates-destr},
and~\ref{thm:partition-candidates} below will make use of the
following lemma.

\begin{lemma}
\label{lem:resistance-candidate-control}
Consider the election $(C,V)$ constructed according to
Construction~\ref{con:resistance-general-candidate-control} from a
$\hittingset$ instance $(B,\mathcal{S},k)$.
\begin{enumerate}
\item If $\mathcal{S}$ has a hitting set $B'$ of size~$k$, then $w$ is
the unique FV winner of election $(B' \cup \{c,d,w\},V)$.

\item Let $D \subseteq B \cup \{d,w\}$.  If $c$ is not a unique FV
  winner of election $(D \cup \{c\},V)$, then there exists a set $B'
  \seq B$ such that
\begin{enumerate}
\item \label{lem:resistance-candidate-control-part-2a}
$D = B' \cup \{d,w\}$, 

\item \label{lem:resistance-candidate-control-part-2b}
$w$ is the unique level~2 FV winner of election $(B' \cup
  \{c,d,w\},V)$, and

\item \label{lem:resistance-candidate-control-part-2c}
$B'$ is a hitting set for $\mathcal{S}$ of size at most~$k$.
\end{enumerate}
\end{enumerate}
\end{lemma}

\begin{proofs}
For the first part, suppose that $B'$ is a hitting set for
$\mathcal{S}$ of size~$k$.  Then there is no level~1 FV winner in
election $(B' \cup \{c,d,w\},V)$, and we have the following level~2
scores:
\begin{eqnarray*}
\scoresublevel{(B' \cup \{c,d,w\},V)}{2}{c}   &  =   & 4n(k+1)+2(m-k)+9,\\
\scoresublevel{(B' \cup \{c,d,w\},V)}{2}{d}   &  =   & 2n(k+1)+2(m+k+1),\\
\scoresublevel{(B' \cup \{c,d,w\},V)}{2}{w}   &  =   & 4n(k+1)+2(m-k)+10,\\
\scoresublevel{(B' \cup \{c,d,w\},V)}{2}{b_j} & \leq & 2n(k+1)+2
\mbox{\hspace{1ex} for all $b_j\in B'$}.
\end{eqnarray*}
It follows that $w$ is the unique level~2 FV winner of election $(B'
\cup \{c,d,w\},V)$.

For the second part, let $D \subseteq B \cup \{d,w\}$. Suppose $c$ is
not a unique FV winner of election $(D \cup \{c\},V)$.
\begin{enumerate}
\item[(\ref{lem:resistance-candidate-control-part-2a})] Other
  than~$c$, only $w$ is approved by a strict majority of voters and
  only $w$ can tie or beat $c$ by the number of approvals in $(D \cup
  \{c\},V)$.  Thus, since $c$ is not a unique FV winner of election
  $(D \cup \{c\},V)$, $w$ is clearly in~$D$.  In $(D\cup \{c\},V)$,
  candidate $w$ has no level~1 strict majority, and candidate $c$ has
  already on level~2 a strict majority.  Thus, $w$ must tie or beat
  $c$ on level~2.  For a contradiction, suppose $d\notin D$.  Then
\[
\scoresublevel{(D \cup \{c\},V)}{2}{c} \geq 4n(k+1)+2m+11.
\]
The overall score of $w$ is
\[
\scoresub{(D\cup \{c\},V)}{w}=4n(k+1)+2m+10,
\]
which contradicts our assumption, that $w$ ties or beats $c$ on
level~2. Thus, $D=B' \cup \{d,w\}$, where $B'\subseteq B$.
   
\item[(\ref{lem:resistance-candidate-control-part-2b})] This part
  follows immediately from
  part~(\ref{lem:resistance-candidate-control-part-2a}).
   
\item[(\ref{lem:resistance-candidate-control-part-2c})] Let $\ell$ be
  the number of sets in $\mathcal{S}$ not hit by $B'$.  We have that
\begin{eqnarray*}
\scoresublevel{(B' \cup \{c,d,w\},V)}{2}{w}
 & = & 4n(k+1)+10+2(m-\| B'\| ), \\
\scoresublevel{(B' \cup \{c,d,w\},V)}{2}{c}
 & = & 2(m-k)+4n(k+1)+9+2(k+1)\ell .
\end{eqnarray*}
From part~(\ref{lem:resistance-candidate-control-part-2a}) we know
that
\[
\scoresublevel{(B' \cup \{c,d,w\},V)}{2}{w} \geq
\scoresublevel{(B' \cup \{c,d,w\},V)}{2}{c},
\]
so
\begin{eqnarray*}
4n(k+1)+10+2(m-\| B'\| )
 & \geq & 2(m-k)+4n(k+1)+9+2(k+1)\ell .
\end{eqnarray*}
The above inequality implies
\[
1 > \frac{1}{2} \geq \| B' \| -k + (k+1)\ell \geq 0,
\]
so $\| B' \| -k + (k+1)\ell = 0$.  Thus $\ell = 0$, and it follows
that $B'$ is a hitting set for $\mathcal{S}$ of size at most~$k$.
\end{enumerate}
This completes the proof of
Lemma~\ref{lem:resistance-candidate-control}.~\end{proofs}

\begin{theorem}
\label{thm:adding-candidates}
Fallback voting is resistant to constructive and destructive control
by adding candidates (both in the limited and the unlimited version of
the problem).
\end{theorem}

\begin{proofs}
Susceptibility holds by Lemma~\ref{lem:susceptible-candidate-control}.
$\np$-hardness follows immediately from
Lemma~\ref{lem:resistance-candidate-control}, via mapping the
$\hittingset$ instance $(B,\mathcal{S},k)$ to the instance
\begin{enumerate}
\item $((\{c,d,w\} \cup B, V), w, k)$ of Constructive Control by
  Adding a Limited Number of Candidates,
\item $((\{c,d,w\} \cup B, V), c, k)$ of Destructive Control by Adding
  a Limited Number of Candidates,
\item $((\{c,d,w\} \cup B, V), w)$ of Constructive Control by
  Adding an Unlimited Number of Candidates, and
\item $((\{c,d,w\} \cup B, V), c)$ of Destructive Control by Adding
  an Unlimited Number of Candidates.
\end{enumerate}
where in each case $c$, $d$, and $w$ are the qualified candidates and
$B$ is the set of spoiler candidates.~\end{proofs}

\begin{theorem}
\label{thm:deleting-candidates-destr}
Fallback voting is resistant to destructive control by deleting
candidates.
\end{theorem}

\begin{proofs}
Susceptibility holds by Lemma~\ref{lem:susceptible-candidate-control}.
To show the problem $\np$-hard, let $(C,V)$ be the election resulting
from a $\hittingset$ instance $(B,\mathcal{S},k)$ according to
Construction~\ref{con:resistance-general-candidate-control}, and let
$c$ be the distinguished candidate.

We claim that $\mathcal{S}$ has a hitting set of size at most~$k$ if
and only if $c$ can be prevented from being a unique FV winner by
deleting at most $m-k$ candidates.

From left to right: Suppose $\mathcal{S}$ has a hitting set $B'$ of
size~$k$. Delete the $m-k$ candidates $B-B'$.  Now, both candidates
$c$ and $w$ have a strict majority on level~2, but
\begin{eqnarray*}
\scoresublevel{(\{c,d,w\} \cup B', V)}{2}{c} & = & 4n(k+1)+2(m-k)+9, \\
\scoresublevel{(\{c,d,w\} \cup B', V)}{2}{w} & = & 4n(k+1)+2(m-k)+10,
\end{eqnarray*}
so $w$ is the unique level~2 FV winner of this election.

From right to left: Suppose that $c$ can be prevented from being a
unique FV winner by deleting at most $m-k$ candidates.  Let $D'
\subseteq B\cup \{ d,w\} $ be the set of deleted candidates (so
$c\notin D'$) and $D = (C-D')-\{ c \}$.  It follows immediately from
Lemma~\ref{lem:resistance-candidate-control} that $D = B'\cup \{ d,w
\}$, where $B'$ is a hitting set for $\mathcal{S}$ of size at
most~$k$.~\end{proofs}

\begin{theorem}
\label{thm:partition-candidates}
Fallback voting is resistant to constructive and destructive control
by partition of candidates and by run-off partition of candidates (for
each in both tie-handling models, TE and~TP).
\end{theorem}

\begin{proofs}
 Susceptibility holds by
 Lemma~\ref{lem:susceptible-candidate-control}, so it remains to show
 $\np$-hardness.  For the constructive cases, map the given
 $\hittingset$ instance $(B,\mathcal{S},k)$ to the election $(C,V)$
 from Construction~\ref{con:resistance-general-candidate-control} with
 distinguished candidate~$w$.

 We claim that $\mathcal{S}$ has a hitting set of size at most~$k$ if
 and only if $w$ can be made a unique FV winner by exerting control
 via any of our four control scenarios (partition of candidates with
 or without run-off, and for each in either tie-handling model, TE and
 TP).

From left to right: Suppose $\mathcal{S}$ has a hitting set
$B'\subseteq B$ of size $k$. Partition the set of candidates into the
two subsets $C_1=B'\cup \{ c,d,w \}$ and $C_2=C-C_1$.  According to
Lemma~\ref{lem:resistance-candidate-control}, $w$ is the unique
level~2 FV winner of subelection $(C_1, V) = (B' \cup \{c,d,w\},V)$.
Note that $w$'s score in the final stage is at least
$2(m-k)+4n(k+1)+9$.  Since (no matter whether we have a run-off or
not, and regardless of the tie-handling rule used) the opponents of
$w$ in the final stage (if there are any opponents at all) each are
candidates from $B$ whose score is at most $2n(k+1)+2$, $w$ is the
only candidate in the final stage with a strict majority of approvals.
Thus, $w$ is the unique FV winner of the resulting election.

From right to left: Suppose $w$ can be made a unique FV winner via
any of our four control scenarios.  Since $c$ is not an FV winner of
the election, there is a subset $D\subseteq B\cup \{ d,w\} $ of
candidates such that $c$ is not a unique FV winner of the election
$(D\cup \{ c\} ,V)$.  By Lemma~\ref{lem:resistance-candidate-control},
there exists a hitting set for $\mathcal{S}$ of size at most~$k$.

For the four destructive cases, we simply change the roles of $c$ and
$w$ in the above
argument.~\end{proofs}

Construction~\ref{con:resistance-general-candidate-control} does not
work for constructive control by deleting candidates in fallback
voting.  By deleting $c$ the chair could make $w$ a unique FV winner,
regardless of whether or not $\mathcal{S}$ has a hitting set of
size~$k$.  The following theorem provides a different construction
that shows resistance in this case as well.

\begin{theorem}
\label{thm:deleting-candidates-constr}
Fallback voting is resistant to constructive control by deleting
candidates.
\end{theorem}

\begin{proofs}
Susceptibility holds by Lemma~\ref{lem:susceptible-candidate-control}.
To show $\np$-hardness, let $(B,\mathcal{S},k)$ be a $\hittingset$
instance with $B=\{b_1,b_2,\ldots ,b_m\}$ a set,
$\mathcal{S}=\{S_1,S_2,\ldots ,S_n\}$ a collection of nonempty subsets
$S_i\subseteq B$, and $k\leq m$ a positive integer.
Define the election $(C,V)$ with candidate set
\[
C = B \cup C' \cup D \cup E \cup \{ w\},
\]
where $C'=\{ c_1, c_2, \ldots ,c_{k+1} \}$, $D=\{d_1,d_2,\ldots
,d_s\}$, $E=\{ e_1, e_2, \ldots , e_n\}$, $w$ is the distinguished
candidate, and the number of candidates in $D$ is $s = \sum _{i=1}^n
s_i$ with $s_i = n+k-\| S_i \|$, so
$s = n^2+kn-\sum _{i=1}^n \| S_i\|$.  Define $V$ to be the
following collection of $2(n+k+1)+1$ voters:
\begin{enumerate}
\item For each $i$, $1\leq i \leq n$, there is one voter of the form:
\[
\begin{array}{c@{\ \ }c@{\ \ }c@{\ \ }c@{\ \ }c}
S_i & D_i & w & | & (B-S_i)\cup (D-D_i)\cup C'\cup E,
\end{array}
\]
where
$D_i=\{
d_{1 + \sum_{j=1}^{i-1} s_j}, \ldots , d_{\sum_{j=1}^{i} s_j} \}$.

\item For each $j$, $1\leq j \leq k+1$, there is one voter of the
  form:
\[
\begin{array}{c@{\ \ }c@{\ \ }c@{\ \ }c@{\ \ }c}
E & (C'-\{c_j\}) & c_j & | & B \cup D\cup \{w\}.
\end{array}
\]
\item There are $k+1$ voters of the form:
\[
\begin{array}{c@{\ \ }c@{\ \ }c}
w & | & B\cup C'\cup D\cup E.
\end{array}
\]
\item There are $n$ voters of the form:
\[
\begin{array}{c@{\ \ }c@{\ \ }c}
C' & | & B\cup D\cup E\cup \{w\}.
\end{array}
\]
\item There is one voter of the form:
\[
\begin{array}{c@{\ \ }c@{\ \ }c@{\ \ }c}
C' & w & | & B\cup D\cup E.
\end{array}
\]
\end{enumerate}

Note that there is no unique FV winner in the above election; the
candidates in $C'$ and $w$ are all level~$n+k+1$ FV winners.

We claim that $\mathcal{S}$ has a hitting set of size $k$ if and only
if $w$ can be made a unique FV winner by deleting at most $k$
candidates.

From left to right: Suppose $\mathcal{S}$ has a hitting set $B'$ of
size~$k$.  Delete the corresponding candidates.  Now, $w$ is the
unique level~$(n+k)$ FV winner of the resulting election.

From right to left: Suppose $w$ can be made a unique FV winner by
deleting at most $k$ candidates.  Since $k+1$ candidates other than
$w$ have a strict majority on level~$n+k+1$ in election $(C,V)$, after
deleting at most $k$ candidates, there is still at least one candidate
other than $w$ with a strict majority of approvals on level~$n+k+1$.
However, since $w$ was made a unique FV winner by deleting at most $k$
candidates, $w$ must be the unique FV winner on a level lower than or
equal to $n+k$.  This is possible only if in all $n$ votes of the
first voter group $w$ moves forward by at least one position.  This,
however, is possible only if $\mathcal{S}$ has a hitting set $B'$ of
size~$k$.~\end{proofs}

\subsection{Voter Control}
\label{sec:results:voter-control}

All vulnerability results in this section follow via
Lemma~\ref{lem:susceptible-voter-control}, showing susceptibility, and
a polynomial-time algorithm for the corresponding control problem.
All resistance results in this section follow via
Lemma~\ref{lem:susceptible-voter-control}, showing susceptibility, and
a reduction from either Hitting Set or
the well-known $\np$-complete problem Exact Cover by
Three-Sets~\cite{gar-joh:b:int}, which is defined as follows:

\begin{desctight}

\item[Name:] Exact Cover by Three-Sets (X3C).

\item[Instance:] A set $B = \{b_1, b_2, \ldots , b_{3m }\}$, $m\geq 1$, and a
collection $\mathcal{S} = \{S_1, S_2, \ldots , S_n\}$ of subsets $S_i
\seq B$ with $\|S_i\| = 3$ for each~$i$, $1 \leq i \leq n$.

\item[Question:] Is there a subcollection $\mathcal{S}' \seq
  \mathcal{S}$ such that each element of $B$ occurs in exactly one set
  in~$\mathcal{S}'$?
\end{desctight}

Our first result for voter control says that fallback voting is
resistant to constructive control by adding voters and to constructive
control by deleting voters.

\begin{theorem}
\label{thm:adding-deleting-voters}
Fallback voting is resistant to constructive control by adding voters
and by deleting voters.
\end{theorem}

\begin{proofs}
Susceptibility holds by Lemma~\ref{lem:susceptible-voter-control} in
both cases.  We first prove $\np$-hardness---and thus resistance---of
Constructive Control by Adding Voters.  Let $(B,\mathcal{S})$ be an
X3C instance, where $B=\{b_1,b_2,\ldots ,b_{3m}\}$ is a set with $m>1$
and $\mathcal{S}=\{S_1,S_2,\ldots ,S_n\}$ is a collection of subsets
$S_i\subseteq B$ with $\|S_i\|=3$ for each $i$, $1 \leq i \leq n$.
(Note that X3C is trivial to solve for $m=1$.)

Define the election $(C,V\cup V')$, where $C=B\cup \{w\} \cup D$ with
$D= \{d_1, \ldots ,d_{n(3m-4)} \}$ is the set of candidates, $w$ is
the distinguished candidate, and $V\cup V'$ is the following collection
of $n+m-2$ voters:\footnote{This construction---just as that
  for SP-AV~\cite{erd-now-rot:j:sp-av}---is based on
  the corresponding construction for approval voting
  \cite{hem-hem-rot:j:destructive-control}.}
\begin{enumerate}
\item $V$ is the collection of $m-2$ registered voters of the form:
\[
 \begin{array}{c@{\ \ }c@{\ \ }c}
B & | & D\cup \{w\}.
\end{array}
\]
\item  $V'$ is the collection of unregistered voters, where for each 
$i$, $1\leq i \leq n$, there is one voter of the form:
\[
\begin{array}{c@{\ \ }c@{\ \ }c@{\ \ }c@{\ \ }c}
D_i & S_i & w & | & (B-S_i) \cup (D-D_i),
\end{array}
\]
where $D_i=\{d_{(i-1)(3m-4)+1}, \ldots , d_{i(3m-4)} \} $.
\end{enumerate}
Since $w$
has no approvals in $(C,V)$, $w$ is not a unique FV winner in $(C,V)$.

We claim that $\mathcal{S}$ has an exact cover for $B$ if and only if
$w$ can be made a unique FV winner by adding at most $m$ voters
from~$V'$.

From left to right: Suppose $\mathcal{S}$ contains an exact cover
for~$B$.  Let $V''$ contain the corresponding voters from~$V'$
(i.e., voters of the form
$\begin{array}{@{\, }c@{\ \ }c@{\ \ }c@{\ \ }c@{\ \ }c@{\, }}
D_i & S_i & w & | & (B-S_i) \cup (D-D_i)
\end{array}$
for each $S_i$ in the exact cover)
and add $V''$ to the election.  It follows that $\scoresub{(C,V\cup
  V'')}{d_j}=1$ for all $d_j\in D$, $\scoresub{(C,V\cup V'')}{b_j}=m-1$
for all $b_j\in B$, and $\scoresub{(C,V\cup V'')}{w}=m$.  Thus, only
$w$ has a strict majority of approvals and so is the unique FV winner
of the election.

From right to left: Let $V''\subseteq V'$ be such that $\|V''\|\leq m$ and
$w$ is the unique winner of election $(C,V\cup V'')$.  Since $w$ must in
particular beat every $b_j \in B$, it follows that
$\|V''\|=m$ and each $b_j\in B$ can gain only one additional point.
Thus, the $m$ voters in $V''$ correspond to an exact cover for~$B$.

Next, we show that FV is resistant to constructive control by deleting
voters.  Let $(B,\mathcal{S})$ be an X3C instance as above.  Define
the election $(C,V)$, where $C=B\cup \{c,w\} \cup D$ is the set of
candidates with $D=\{ d_1, d_2, \ldots ,d_{3nm}\}$, $w$ is the
distinguished candidate, and $V$ is the following collection of
$2n+m-1$ voters:
\begin{enumerate}
\item For each $i$, $1\leq i \leq n$, there is
one voter of the form:
\[
\begin{array}{c@{\ \ }c@{\ \ }c@{\ \ }c}
S_i & c & | & (B-S_i)\cup D \cup \{w\}.
\end{array}
\]
\item For each $i$, $1\leq i \leq n$, there is one 
voter of the form:
\[
\begin{array}{c@{\ \ }c@{\ \ }c@{\ \ }c@{\ \ }c}
B_i & D_i & w & | & (B-B_i)\cup \{ c\} \cup (D-D_i),
\end{array}
\]
where, letting
$\ell_j = \| \{S_i \in \mathcal{S} \condition b_j\in S_i \}\|$
for each~$j$, $1\leq j \leq 3m$,
$B_i = \{b_j\in B \condition i\leq n-\ell _j\}$ and
$D_i = \{ d_{(i-1)3m+1}, \ldots , d_{3im-\| B_i \|} \}$.

Note that $D_i=\emptyset$ if $\| B_i \| =3m$.  Note also that $w$ is
always ranked on the $(3m+1)^{st}$ place.
\item There are $m-1$ voters of the form:
\[
\begin{array}{c@{\ \ }c@{\ \ }c}
c & | & B\cup D \cup \{w\}.
\end{array}
\]
\end{enumerate}

Note that $\scoreof{w}=\scoreof{b_i}=n$ for all $i$, $1\leq i \leq
3m$, $\scoreof{d_j}=1$ or $\scoreof{d_j}=0$ for all $d_j\in D$, and
$\scoreof{c}= n+m-1$, and since only $c$ has a strict majority (which
is reached on level~$4$), $c$ is the unique level~$4$ FV winner of the
election.

We claim that $\mathcal{S}$ has an exact cover for $B$ if and only if
$w$ can be made a unique FV winner by deleting at most $m$ voters.

From left to right: Suppose $\mathcal{S}$ contains an exact cover for
$B$. By deleting the corresponding voters from the first voter group,
we have the following scores: $\scoreof{w}=n$,
$\scoreof{b_i}=\scoreof{c}=n-1$ for all $i$, $1\leq i \leq 3m$, and
$\scoreof{d_j}=1$
or $\scoreof{d_j}=0$ for all $d_j\in D$. Since there are now $2n-1$
voters in the election, only candidate $w$ has a strict majority, so
$w$ is the unique FV winner of the election.

From right to left: Suppose $w$ can be made a unique FV winner by
deleting at most $m$ voters.  Since $w$'s approvals are all on the
$(3m+1)^{st}$ position, neither $c$ nor any of the $b_i$ can have a
strict majority on any of the previous levels.  In particular,
candidate $c$ must have lost exactly $m$ points after deletion, and
this is possible only if the $m$ deleted voters are all from the first
or third voter group.  On the other hand, each $b_i\in B$ must have
lost at least one point after deletion, and this is possible only if
exactly $m$ voters were deleted from the first voter group.  These $m$
voters correspond to an exact cover for~$B$.~\end{proofs}

In contrast to the constructive voter-control cases of
Theorem~\ref{thm:adding-deleting-voters}, fallback voting is
vulnerable to destructive control by adding voters and to destructive
control by deleting voters.  In fact, the proof of
Theorem~\ref{thm:destructive-adding-deleting-voters} shows something
slightly stronger: FV is what
\cite{hem-hem-rot:j:destructive-control} call ``certifiably
vulnerable'' to these two destructive voter-control cases, i.e., the
algorithm we present in this proof for destructive control by adding
voters even computes a successful control
action if one exists (instead of only solving the corresponding
decision problem).\footnote{And the same holds for the algorithm
  showing that FV is vulnerable to destructive control by deleting
  voters, which is not presented here due to space.}

\begin{theorem}
\label{thm:destructive-adding-deleting-voters}
Fallback voting is vulnerable to destructive control by adding voters
and by deleting voters.
\end{theorem}

\begin{proofs}
  Susceptibility holds by Lemma~\ref{lem:susceptible-voter-control} in
  both cases.  We present a polynomial-time algorithm for solving the
  destructive control by adding voters case.  We will make use of the
  following notation.  Given an election $(C,V)$, let $\mathit{maj}(V)
  = \left\lfloor \nicefrac{\|V\|}{2} \right\rfloor +1$ and define the
  deficit of candidate $d \in C$ for reaching a strict majority in
  $(C,V)$ on level~$i$, $1 \leq i \leq \|C\|$,
  by
\[
\mathit{def}_{(C,V)}^i(d) =
\mathit{maj}(V)-\scoresublevel{(C,V)}{i}{d}.
\]

The input to our algorithm is an election $(C,V\cup V')$ (where $C$ is
the set of candidates, $V$ is the collection of
registered voters, and $V'$ is the collection of unregistered voters),
a distinguished candidate $c\in C$, and an integer $\ell$ (the number
of voters allowed to be added).  The algorithm either outputs a subset
$V''\subseteq V'$, $\|V''\| \leq \ell$, that describes a successful
control action (if any exists), or indicates that control is
impossible for this input.

We give a high-level description of the algorithm.  We assume that $c$
is initially the unique FV winner of election $(C,V)$; otherwise, the
algorithm simply outputs $V'' = \emptyset$ and halts,
since there is no need to add any voters from~$V'$.

Let $n = \max_{v \in V \cup V'} \|S_v\|$.  Clearly, $n \leq \|C\|$.
The algorithm proceeds in at most $n+1$
stages, where the last stage is the \emph{approval stage} and checks
whether $c$ can be dethroned as a unique FV winner by approval score
via adding at most $\ell$ voters from~$V'$, and all preceding stages
are \emph{majority stages} that check whether a candidate $d \neq c$ can
tie or beat $c$ on level~$i$ via adding at most $\ell$ voters
from~$V'$.  Since the first majority stage is slightly different from
the subsequent majority stages, we describe both cases separately.

\paragraph{Majority Stage~1:} For each candidate $d\in C-\{ c\}$,
check whether $d$ can tie or beat $c$ on the first level via adding at
most $\ell$ voters from~$V'$.  To this end, first check whether
\begin{eqnarray}
\label{equ:check-1}
\mathit{def}_{(C,V)}^1(d)    & \leq & \frac{\ell}{2}; \\
\label{equ:check-2}
\scoresublevel{(C,V)}{1}{d} & \geq & \scoresublevel{(C,V)}{1}{c} - \ell.
\end{eqnarray}
If (\ref{equ:check-1}) or (\ref{equ:check-2}) fails to hold, this $d$
is hopeless, so go to the next candidate (or to the next stage if all
other candidates have already been checked in this stage).
If (\ref{equ:check-1}) and (\ref{equ:check-2}) hold, find a set $V'_d
\subseteq V'$ of largest cardinality such that $\|V'_d\| \leq \ell$
and all voters in $V'_d$ approve of $d$ on the first level but
disapprove of $c$ on the first level.  Check whether
\begin{eqnarray}
\label{equ:check-3}
\scoresublevel{(C,V\cup V'_d)}{1}{d} \geq 
\scoresublevel{(C,V\cup V'_d)}{1}{c}.
\end{eqnarray}
If (\ref{equ:check-3}) fails to hold, this $d$ is hopeless, so go to
the next candidate (or to the next stage if all other candidates have
already been checked in this stage).  If (\ref{equ:check-3}) holds,
check whether $d$ has a strict majority in $(C,V\cup V'_d)$
on the first level, and if so,
output $V'' = V_d'$ and halt.

\paragraph{Majority Stage~\textit{i}, 1 $<$ \textit{i} $\leq$
  \textit{n}:} This stage is entered only if it was not possible to
find a successful control action in majority stages $1, \ldots , i-1$.
For each candidate $d\in C-\{ c\}$, check whether $d$ can tie or beat
$c$ up to the $i$th level via adding at most $\ell$ voters from~$V'$.  To
this end, first check whether
\begin{eqnarray}
\label{equ:i-check-1}
\mathit{def}_{(C,V)}^i(d)    & \leq & \frac{\ell}{2}; \\
\label{equ:i-check-2}
\scoresublevel{(C,V)}{i}{d} & \geq & \scoresublevel{(C,V)}{i}{c} - \ell.
\end{eqnarray}
If (\ref{equ:i-check-1}) or (\ref{equ:i-check-2}) fails to hold, this
$d$ is hopeless, so go to the next candidate (or to the next stage if
all other candidates have already been checked in this stage).  If
(\ref{equ:i-check-1}) and (\ref{equ:i-check-2}) hold, find a set $V'_d
\subseteq V'$ of largest cardinality such that $\|V'_d\| \leq \ell$
and all voters in $V'_d$ approve of $d$ up to the $i$th level but
disapprove of $c$ up to the $i$th level.  Check whether
\begin{eqnarray}
\label{equ:i-check-3}
\scoresublevel{(C,V\cup V'_d)}{i}{d} \geq 
\scoresublevel{(C,V\cup V'_d)}{i}{c}
\end{eqnarray}
If (\ref{equ:i-check-3}) fails to hold, this $d$ is hopeless, so go to
the next candidate (or to the next stage if all other candidates have
already been checked in this stage).  If (\ref{equ:i-check-3}) holds,
check whether $d$ has a strict majority in $(C,V\cup V'_d)$
on the $i$th level, and if so,
check whether
\begin{eqnarray}
\label{equ:i-check-4}
\scoresublevel{(C,V\cup V'_d)}{i-1}{c} \geq
\mathit{maj}(V\cup V'_d).
\end{eqnarray}
If (\ref{equ:i-check-4}) fails to hold, output $V'' = V_d'$ and halt.
Otherwise (i.e., if (\ref{equ:i-check-4}) holds), though $d$ might be able
to dethrone $c$ by adding $V'_d$ on the $i$th level, this was not early
enough, since $c$ has already won at a previous level.
In that case, find a largest set $V'_{cd} \subseteq V'$ such that
\begin{enumerate}
\item $\|V'_d \cup V'_{cd}\| \leq \ell$,
\item all voters in $V'_{cd}$ approve of both $c$ and $d$ up to the
  $i$th level, and
\item the voters in $V'_{cd}$ are chosen such that $c$ is approved of
  as late as possible by them (i.e., at levels with a largest possible
number, where ties may be broken arbitrarily).
\end{enumerate}
Now, check whether
\begin{eqnarray}
\label{equ:i-check-5}
\scoresublevel{(C,V\cup V'_d\cup V'_{cd})}{i-1}{c} \geq
\mathit{maj}(V\cup V'_d\cup V'_{cd}).
\end{eqnarray}
If (\ref{equ:i-check-5}) holds, then this $d$ is hopeless, so go to the
next candidate (or to the next stage if all other candidates have
already been checked in this stage).  Otherwise (i.e., if
(\ref{equ:i-check-5}) fails to hold), check whether $\|V'_{cd}\| \geq
\mathit{def}_{(C,V \cup V'_d)}^i(d)$.  If so (i.e., $d$ has now a strict
majority on level~$i$), output $V'' = V_d' \cup V'_{cd}$ and halt.
Note that, by choice of $V'_{cd}$, (\ref{equ:i-check-3}) implies that
\[
\scoresublevel{(C,V\cup V'_d\cup V'_{cd})}{i}{d} \geq 
\scoresublevel{(C,V\cup V'_d\cup V'_{cd})}{i}{c}.
\]
Thus, in $(C,V\cup V'_d\cup V'_{cd})$,
$d$ ties or beats $c$ and has a strict majority on the $i$th level
(and now, we are sure that $d$ was not too late).
Otherwise (i.e., if $\|V'_{cd}\| < \mathit{def}_{(C,V \cup V'_d)}^i(d)$),
this $d$ is hopeless, so go to the next candidate (or stage).

\paragraph{Approval Stage:} This stage is entered only if it was not possible to
find a successful control action in majority stages $1, \ldots , n$.
First, check if
\begin{eqnarray}
\label{equ:i-check-6}
\scoresub{(C,V)}{c} < \left\lfloor \frac{\| V\| +
    \ell}{2} \right\rfloor +1.
\end{eqnarray}
If (\ref{equ:i-check-6}) fails to hold, output ``control impossible'' and
halt, since we have found no candidate in the majority stages who could tie
or beat $c$ and have a strict majority when adding at most $\ell$ voters
from~$V'$, so adding any choice of at most $\ell$ voters from $V'$ would
$c$ still leave a strict majority.  If (\ref{equ:i-check-6}) holds,
looping over all candidates $d\in C-\{ c\}$, check whether
there are $\scoresub{(C,V)}{c} - \scoresub{(C,V)}{d} \leq \ell$
voters in $V'$ who approve of $d$ and
disapprove of~$c$.  If this is not the case, move on to the next
candidate, since $d$ could never catch up on $c$ via adding at most $\ell$
voters from $V'$.  If it is the case for some $d\in C-\{ c\}$, however,
add this set of voters (call it $V_d'$) and check whether
\begin{eqnarray}
\label{equ:i-check-7}
\scoresub{(C,V \cup V_d')}{c} < \mathit{maj}(V\cup V_d').
\end{eqnarray}
If (\ref{equ:i-check-7}) holds, output $V'' = V_d'$ and halt.
Otherwise (i.e., if (\ref{equ:i-check-7}) fails to hold), check
whether
\begin{eqnarray}
\label{equ:i-check-8}
\ell - \|V_d'\| & \geq & \|V_{\emptyset}' \| \\
                & \geq & 2\left(\scoresub{(C,V \cup V_d')}{c}
                         - \frac{\| V \cup V_d'\|}{2}\right), \nonumber 
\end{eqnarray}
where $V_{\emptyset}'$ is contained in the set of voters in $V'$
who disapprove of
both candidates $c$ and~$d$.  If (\ref{equ:i-check-8}) does not hold,
move on to the next candidate, since after adding these voters $c$
would still have a strict majority.  Otherwise (i.e., if
(\ref{equ:i-check-8}) holds), add exactly $2\left(\scoresub{(C,V \cup
  V_d')}{c} - \nicefrac{\| V \cup V_d'\|}{2}\right)$ voters from
$V_{\emptyset}'$ (denoted by $V_{\emptyset,+}'$).  Output $V'' = V_d'
\cup V_{\emptyset,+}'$ and halt.

If we have entered the approval stage (because we were not successful
in any of the majority stages), but couldn't find any candidate here
who was able to dethrone $c$ by adding at most $\ell$ voters
from~$V'$, we output ``control impossible'' and halt.

The correctness of the algorithm follows from the remarks made above.
Crucially, note that the algorithm proceeds in the ``safest way
possible'': If there is any successful control action at all then our
algorithm finds some successful control action.  It is also easy to
see that this algorithm runs in polynomial time.  (Note that we didn't
optimize it in terms of running time; rather, we described it in a way
to make it easier to check its correctness.)

The deleting-voters case follows by a similar algorithm (and is
omitted here due to space).
\end{proofs}

\OMIT{
\begin{proofs}
  We first present a polynomial-time algorithm for solving the
  destructive control by adding voters case.  We will make use of the
  following notation.  Let $\mathit{def}_i(d)$defined as
\[
\mathit{def}_i(d) = \mathit{maj}(V)-\scoresublevel{(C,V)}{i}{d}
\]
be the deficit of candidate $d$ on level~$i$, where
$\mathit{maj}(V)= \bigg\lfloor \nicefrac{\| V \|}{2} \bigg\rfloor +1$. 
 
The input to our algorithm is an election $(C,V\cup V')$,
where $C$ is the set of candidates,
$V$ is the collection of registered voters and 
$V'$ is the collection of unregistered voters, a 
distinguished candidate $c\in C$, and an integer $\ell $, 
which is the number of voters allowed to be added. 
The output is a subset $V''\subseteq V'$.
The algorithm checks step by step, whether a candidate
$d\in C-\{ c\}$ could tie or beat $c$ in a head-to-head contest on 
level~$i$ by strict majority, where $1\leq i \leq n$
(i.e., $c$ gets no strict majority on a level earlier than $i$). 
If there is no such candidate, then the algorithm checks
whether a candidate  $d\in C-\{ c\}$ could tie or beat $c$ by 
score (i.e., after adding the voters in $V''$, both $c$ and $d$ 
have no majority). The algorithm works as follows:

\begin{enumerate}
 \item {\bf Checking the trivial cases:} If $c$ is already
not the unique FV winner of the election, then output
$\emptyset $ since there is no need to add any voters from~$V'$. 

 \item {\bf Loop 1:} For each level~$i$, $1\leq i \leq n$, and 
  each candidate $d\in C-\{ c\}$, 
  \begin{itemize}
   \item check if $\mathit{def}_i(d)\leq \nicefrac{\ell }{2}$, and if 
   $\scoresublevel{(C,V)}{i}{d} \geq \scoresublevel{(C,V)}{i}{c} - \ell$.
   If no, move to the next level. If yes, 
   \item add the set of all voters (maximally $\ell $) $V_d' \subseteq V'$ from $V'$, 
   who approve of $d$ on the first $i$ levels, and disapprove of $c$ on the 
   first $i$ levels. Check if $c$ has strict majority on level $i-1$. If not, and $d$ 
   already has strict majority on level~$i$, and $d$ ties or beats $c$ there, then 
   output $V_d'$, and halt.
   
   If $c$ has strict majority on level $i-1$, or $d$ has still no strict majority on
   level $i$, or if $d$ has strict majority on level $i$ but does not tie or beat
   $c$ on that level, check whether  $\scoresublevel{(C,V\cup V_d')}{i}{d} \geq 
   \scoresublevel{(C,V\cup V_d')}{i}{c}$. If no, then move to the next level, 
   since from now on we can only add voters who approve of both candidates 
   $c$ and $d$ on the first $i$ levels, thus the score of $c$ on level~$i$ will 
   be always higher than the score of $d$ on level~$i$. If yes, 

   \item add $\mathit{def}_i(d)- \|V_d' \|$ voters (if there exists as many voters with these
   properties), who approve of both $c$ and $d$ on the first $i$ levels, say 
   $V_{cd}' \subseteq V'$. Choose these voters in a way, that $c$ should get his or her 
   approvals as late as possible (i.e., on the deepest levels as possible).
   If  $\scoresublevel{(C,V\cup V_d' \cup V_{cd}')}{i}{d} \geq 
   \scoresublevel{(C,V\cup V_d'\cup V_{cd}')}{i}{c}$, then check whether $c$ has already strict 
   majority on level~$i-1$. If yes, move to the next level, if no, output $V_d' \cup V_{cd}'$, 
   and halt.
  \end{itemize}

  If no candidate $d$ was found in Loop 1, who could tie or beat $c$ on any level~$i$, then 
  move to Loop 2.

 \item {\bf Loop 2:} 
   \begin{itemize}
    \item Check if $\scoresub{(C,V)}{c} < \big\lfloor \nicefrac{\| V\| + \ell}{2} \big\rfloor +1$.
    If no, output ``control impossible'', since we have found no candidate in Loop 1 who 
    could tie or beat $c$ by strict majority, and in this case, by adding at most $\ell $ voters,
    $c$ would still have strict majority.
    If yes, 

    \item check for each candidate $d\in C-\{ c\}$, if there are 
    $\scoresub{(C,V)}{c} - \scoresub{(C,V)}{d} $ voters in $V'$,
    who approve of $d$ and disapprove of $c$. If no, move to the next candidate, since
    $d$ could never close in on $c$. If yes, add this set of voters denoted by $V_d'$ and 
    check whether $\scoresub{(C,V \cup V_d')}{c} < \mathit{maj}(V\cup V_d')$. 
    If yes, output $V_d'$ and halt. If not, check if
    $\| V_{\emptyset}' \| \geq 2(\scoresub{(C,V \cup V_d')}{c} - \nicefrac{\| V \cup V_d'\|}{2})$, 
    where $V_{\emptyset}'$ is the
    set of voters in $V'$, who disapprove of both candidates $c$ and $d$. If the above 
    inequality does not hold, move to the next candidate, since after adding these voters $c$
    would still have strict majority. If the inequality holds, add
    exactly $2(\scoresub{(C,V \cup V_d')}{c} - \nicefrac{\| V \cup V_d'\|}{2})$ voters from  
    $V_{\emptyset}'$
    (denoted by  $V_{\emptyset,+}'$). Thus, output $V_d' \cup V_{\emptyset,+}'$ and halt.
   \end{itemize}

 \item {\bf Termination:} If in no iteration of either Loop 1 or Loop 2 an appropriate 
 subset of $V'$ was found, then output ``control impossible'' and halt.
\end{enumerate}

Finally, we prove that FV is vulnerable to destructive control by deleting voters. 
The algorithm works like the algorithm above with minor changes:

The input to our algorithm is an election $(C,V)$,
where $C$ is the set of candidates with $\| C\| =n$ and
$V$ is the collection of voters, a 
distinguished candidate $c\in C$, and an integer $\ell $, 
which is the number of voters allowed to be deleted. 
The output is a subset $V'\subseteq V$, with $\| V' \| \leq \ell$.
The algorithm checks step by step, whether a candidate
$d\in C-\{ c\}$ could tie or beat $c$ in a head-to-head contest on 
level~$i$ by strict majority, where $1\leq i \leq n$
(i.e., $c$ gets no strict majority on a level earlier than $i$). 
If there is no such candidate, then the algorithm checks
whether a candidate  $d\in C-\{ c\}$ could tie or beat $c$ by 
score (i.e., after deleting the voters, both $c$ and $d$ have no 
majority). The algorithm works as follows:

\begin{enumerate}
 \item {\bf Checking the trivial cases:} If $c$ is already
not the unique FV winner of the election, then output
$\emptyset $ since there is no need to delete any voters from~$V$. 

 \item {\bf Loop 1:} For each level~$i$, $1\leq i \leq n$, and 
  each candidate $d\in C-\{ c\}$, 
  \begin{itemize}
   \item check if $\mathit{def}_i(d)\leq \nicefrac{\ell }{2}$, and if 
   $\scoresublevel{(C,V)}{i}{d} \geq \scoresublevel{(C,V)}{i}{c} - \ell$.
   If no, move to the next level. If yes, 
   \item delete the set of all voters (maximally $\ell $) $V_d \subseteq V$ from $V$, 
   who approve of $c$ on the first $i$ levels, and disapprove of $d$ on the 
   first $i$ levels. 
   Check if $c$ has strict majority on level $i-1$. If not, and $d$ 
   already has strict majority on level~$i$, and $d$ ties or beats $c$ there, then 
   output $V_d$, and halt.
   
   If $c$ has strict majority on level $i-1$, or $d$ has still no strict majority on
   level $i$, or if $d$ has strict majority on level $i$ but does not tie or beat
   $c$ on that level, check whether  $\scoresublevel{(C,V - V_d)}{i}{d} \geq 
   \scoresublevel{(C,V - V_d)}{i}{c}$. If no, then move to the next level, 
   since from now on we can only delete voters who approve of both candidates 
   $c$ and $d$ on the first $i$ levels, thus the score of $c$ on level~$i$ will 
   be always higher than the score of $d$ on level~$i$. If yes, 

   \item delete $\mathit{def}_i(d)- \|V_d \|$ voters (if there exists as many voters with these
   properties), who approve of both $c$ and $d$ on the first $i$ levels, say 
   $V_{cd} \subseteq V$. Choose these voters in a way, that $c$ should get his or her 
   approvals as early as possible (i.e., on the highest levels as possible).
   If  $\scoresublevel{(C,V - (V_d \cup V_{cd}))}{i}{d} \geq 
   \scoresublevel{(C,V - (V_d\cup V_{cd}))}{i}{c}$, then check whether $c$ has already strict 
   majority on level~$i-1$. If yes, move to the next level, if no, output $V_d \cup V_{cd}$, 
   and halt.
  \end{itemize}

  If no candidate $d$ was found in Loop 1, who could tie or beat $c$ on any level~$i$, then 
  move to Loop 2.

 \item {\bf Loop 2:} 
   \begin{itemize}
    \item Check if $\scoresub{(C,V)}{c} < \big\lfloor \nicefrac{\| V\| + \ell}{2} \big\rfloor +1$.
    If no, output ``control impossible'', since we have found no candidate in Loop 1 who 
    could tie or beat $c$ by strict majority, and in this case, by deleting at most $\ell $ voters,
    $c$ would still have strict majority.
    If yes, 

    \item check for each candidate $d\in C-\{ c\}$, if 
     $\scoresub{(C,V)}{c} - \scoresub{(C,V)}{d} \leq \ell$. If there is no such candidate,
    output ``control impossible'', since even if we could delete $\ell$ voters who only
    approve of $c$ and disapprove of all other candidates, $c$ would still have the 
    highest score among all candidates, thus $c$ would win the election.
    If there is a candidate $d$, such that $\scoresub{(C,V)}{c} - \scoresub{(C,V)}{d} \leq \ell$,
    delete all voters, who approve of $c$ and disapprove of $d$. Denote this set by 
    $V_c\subseteq V$. Check if 
    $\scoresub{(C,V - V_c)}{c} < \mathit{maj}(V-V_c)$. If yes, 
    output $V_c$ and halt. If not, delete all voters (but at most $\ell -\| V_c \|$ voters)
    who approve of both candidates $c$ and $d$. Denote this set by $V_{cd}\subseteq V-V_c$.
    Output $V_c \cup V_{cd}$ and halt.
   \end{itemize}

 \item {\bf Termination:} If in no iteration of either Loop 1 or Loop 2 an appropriate 
 subset of $W$ was found, then output ``control impossible'' and halt.
\end{enumerate}
\end{proofs}
}

\begin{theorem}
 \label{thm:constructive-partition-voters-TE}
Fallback voting is resistant to constructive control by 
partition of voters in model~TE. 
\end{theorem}

\begin{proofs}
Susceptibility holds by Lemma~\ref{lem:susceptible-voter-control}.
To prove $\np$-hardness, we reduce X3C to our control problem.
Let $(B,\mathcal{S})$ be an X3C instance with 
$B=\{b_1,b_2,\ldots ,b_{3m}\}$, $m\geq 1$, and a collection
$\mathcal{S}=\{S_1,S_2,\ldots ,S_n\}$ 
of subsets $S_i\subseteq B$ with $\|S_i\|=3$ for each $i$,
$1 \leq i \leq n$.
Our construction, like the corresponding one for
SP-AV~\cite{erd-now-rot:j:sp-av}, is based on
the corresponding construction for approval voting
\cite{hem-hem-rot:j:destructive-control}.
Define the election $(C,V)$, where 
$C=B\cup \{c,x,y,w \} \cup Z$ is the set of candidates, 
$Z=\{ z_1, z_2, \ldots ,z_n\}$, and $w$ is the
distinguished candidate. Define the value $\ell_j = \| \{S_i \in
\mathcal{S} \condition b_j\in S_i \}\|$ for each~$j$, $1\leq j \leq
3m$.  Let $V$ consist of the following $4n+2m-1$ voters:
\begin{enumerate}
\item For each $i$, $1\leq i \leq n$, there is one voter of the form:
\[
 \begin{array}{c@{\ \ }c@{\ \ }c@{\ \ }c}
c & S_i & | & (B-S_i) \cup \{ x,y,w\} \cup Z.
\end{array}
\]
\item For each $i$, $1\leq i \leq n$, there is one voter of the form:
\[
 \begin{array}{c@{\ \ }c@{\ \ }c@{\ \ }c@{\ \ }c@{\ \ }c}
(Z-\{ z_i\}) & B_i & w & | & (B-B_i) & \{ c,x,y,z_i\},
\end{array}
\]
where $B_i=\{ b_j\in B \condition i\leq n-\ell_j\}$.
\item For each $i$, $1\leq i \leq n$, there is one voter of the form:
\[
 \begin{array}{c@{\ \ }c@{\ \ }c@{\ \ }c}
c & z_i & | & B \cup \{ x,y,w\} \cup (Z-\{ z_i\}).
\end{array}
\]
\item There are $n+m$ voters of the form:
\[
 \begin{array}{c@{\ \ }c@{\ \ }c}
x & | & B \cup \{ c,y,w\} \cup Z.
\end{array}
\]
\item There are $m-1$ voters of the form:
\[
 \begin{array}{c@{\ \ }c@{\ \ }c}
y & | & B \cup \{ c,x,w\} \cup Z.
\end{array}
\]
\end{enumerate}

Note that for each $i \in \{1, \ldots , n\}$ and for each $j \in \{1,
\ldots , 3m\}$, we have
$\scoresub{(C,V)}{b_j}=\scoresub{(C,V)}{z_i}=\scoresub{(C,V)}{w}=n$,
$\scoresub{(C,V)}{c}=2n$, $\scoresub{(C,V)}{x}=n+m$, and
$\scoresub{(C,V)}{y}=m-1$.  Thus, there is no candidate with a strict
majority on any level in election $(C,V)$ and, in particular,
candidate $w$ is not a unique FV winner.

We claim that $\mathcal{S}$ has an exact cover for $B$ if and
only if $w$ can be made a unique FV winner of the resulting
election by partition of voters in model~TE.

From left to right: Suppose $\mathcal{S}$ contains an exact cover
$\mathcal{S}'$ for $B$. Partition $V$ in the following way. Let $V_1$
consist of:
\begin{itemize}
\item the $m$ voters of the first group that correspond to the exact
  cover (i.e., those $m$ voters of the form
$\begin{array}{@{\, }c@{\ \ }c@{\ \ }c@{\ \ }c@{\, }}
c & S_i & | & (B-S_i) \cup \{ x,y,w\} \cup Z
\end{array}$
for which $S_i \in \mathcal{S}'$),
\item the $n$ voters of the third group (who approve of $c$
  and~$z_i$), and
\item the $n+m$ voters of the fourth group (who approve of only
$x$).
\end{itemize}
Let $V_2=V-V_1$. In subelection $(C,V_1)$, no candidate has a strict
majority on any level, and $c$ and $x$ tie for first place on the
first level, both with
score $n+m = \nicefrac{\|V_1\|}{2}$.  Thus, there is no candidate
proceeding forward to the final round.  In subelection $(C,V_2)$, only
candidate $w$ has a strict majority, so $w$ is the only participant in
the final round of the election and thus has been made a unique FV
winner by this partition of voters.

From right to left: Suppose that $w$ can be made a unique FV winner by
exerting control by partition of voters in model~{TE}. We can argue for
FV as \cite{erd-now-rot:j:sp-av} do for SP-AV (see also
\cite{hem-hem-rot:j:destructive-control}): Let $(V_1,V_2)$ be such a
successful partition.  Since we are in model~TE, $w$ has to be the
unique winner of one of the subelections, say of $(C,V_1)$. Each voter
of the form
$\begin{array}{@{\, }c@{\ \ }c@{\ \ }c@{\ \ }c@{\, }}
  c & z_i & | & B \cup \{ x,y,w\} \cup (Z-\{ z_i\})
\end{array}$ 
has to be in $V_2$, for otherwise there would be a candidate $z_i\in Z$ with
$\scoresub{(C,V_1)}{z_i} =\scoresub{(C,V_1)}{w} = n$, and $z_i$ would
get his or her approvals on an earlier level than~$w$. Thus, $w$ would
not be the unique winner of subelection $(C,V_1)$.
On the other hand, there can be only $m$ voters of the form
$\begin{array}{@{\, }c@{\ \ }c@{\ \ }c@{\ \ }c@{\, }}
c & S_i & | & (B-S_i) \cup \{ x,y,w\} \cup Z
\end{array}$ 
in~$V_2$, for otherwise $c$ would have the highest score in subelection
$(C,V_2)$, namely $\scoresub{(C,V_2)}{c} > n+m$, and $c$ would reach
this score already on level~$1$. Thus, $c$ would be the unique winner of
subelection $(C,V_2)$, and would also beat $w$ in the run-off because
none of the candidates $c$ and $w$ would have a strict majority in election
$(\{c,w\}, V)$, but $c$ would beat $w$ by approval score.
So there are at most $m$ voters of the form
$\begin{array}{@{\, }c@{\ \ }c@{\ \ }c@{\ \ }c@{\, }}
  c & S_i & | & (B-S_i) \cup \{ x,y,w\} \cup Z
\end{array}$
in~$V_2$.  However, there must be exactly $m$ such voters in $V_2$ and
these $m$ voters correspond to an exact cover for~$B$,
since otherwise there would be a candidate $b_j\in B$ that has
the same score in subelection $(C,V_1)$ as~$w$, namely $n$ points, and
$b_j$ would get his or her approvals on an earlier level than~$w$,
contradicting the assumption that $w$ is the unique FV winner of
subelection $(C,V_1)$.~\end{proofs}

Finally, we turn to control by partition of voters in model~{TP}.
We start with the constructive case.

\begin{theorem}
 \label{thm:constructive-partition-voters-TP}
Fallback voting is resistant to constructive control by 
partition of voters in model~TP. 
\end{theorem}

\begin{proofs}
Susceptibility holds by Lemma~\ref{lem:susceptible-voter-control}.
The proof of resistance is based on the construction of
\cite[Thm.~3.14]{erd-now-rot:j:sp-av}.
Let $(B,\mathcal{S})$ be an X3C instance with 
$B=\{b_1,b_2,\ldots ,b_{3m}\}$, $m\geq 1$, and a collection
$\mathcal{S}=\{S_1,S_2,\ldots ,S_n\}$ 
of subsets $S_i\subseteq B$ with $\|S_i\|=3$ for each $i$,
$1 \leq i \leq n$, and let $n>m+1$.
Define the election $(C,V)$, where 
$C=B\cup F \cup Z \cup \{w,x,y \}$ is the set of candidates, where 
$F = \{f_1,f_2, \ldots, f_{n+m+1}\}$,
$Z=\{z_1, z_2,\ldots ,z_n\}$, and 
$w$ is the distinguished candidate. 
Define the value $\ell_j = \| \{S_i \in
\mathcal{S} \condition b_j\in S_i \}\|$ for each~$j$, $1\leq j \leq
3m$.  Let $V$ consist of the following $6n+2m+2$ voters:
\begin{enumerate}
\item For each $i$, $1\leq i \leq n$, there is one voter of the form:
\[
 \begin{array}{c@{\ \ }c@{\ \ }c@{\ \ }c}
 y & S_i & | & (B-S_i) \cup F \cup Z \cup \{ w,x\}.
\end{array}
\]
\item For each $i$, $1\leq i \leq n$, there is one voter of the form:
\[
 \begin{array}{c@{\ \ }c@{\ \ }c@{\ \ }c}
 y & z_i & | & B \cup F \cup (Z-\{z_i\}) \cup \{ w,x\}.
\end{array}
\]
\item For each $i$, $1\leq i \leq n$, there is one voter of the form:
\[
 \begin{array}{c@{\ \ }c@{\ \ }c@{\ \ }c@{\ \ }c}
 (Z-\{z_i\}) & B_i & w & | & (B-B_i) \cup F \cup \{ x,y,z_i \},
\end{array}
\]
where $B_i=\{ b_j\in B \condition i\leq n-\ell_j\}$.
\item There are $n$ voters of the form:
\[
 \begin{array}{c@{\ \ }c@{\ \ }c@{\ \ }c@{\ \ }c}
 Z & B & w & | & F \cup \{ x,y \}.
\end{array}
\]
\item There are $n+m+1$ voters of the form:
\[
 \begin{array}{c@{\ \ }c@{\ \ }c}
x & | & B \cup F \cup Z \cup \{ w,y\}.
\end{array}
\]
\item For each $k$, $1\leq k \leq n+m+1$, there is one voter of the form:
\[
 \begin{array}{c@{\ \ }c@{\ \ }c}
f_k & | & B \cup (F-\{f_k\}) \cup Z \cup \{ w,x,y\}.
\end{array}
\]
\end{enumerate}

The overall scores of the candidates in election $(C,V)$ can 
be seen in Table~\ref{tab:scores1}.

\begin{table}[h]
 \centering
 \begin{tabular}{|c||c|c|c|c|c|c|}
  \hline
                         & $w$  & $x$     & $y$  & $b_{j}$ & $f_k$ & $z_i$ \\ \hline\hline
  $\tabscoresub{(C,V)}$  & $2n$ & $n+m+1$ & $2n$ & $2n$    & $1$   & $2n$  \\ \hline
 \end{tabular}
\caption{Overall scores in $(C,V)$.}	\label{tab:scores1}
\end{table}

Since the strict majority threshold for $V$ is $3n+m+2$, there is no candidate 
with a strict majority on any level in election $(C,V)$ and, in particular,
since $\scoresub{(C,V)}{y}=\scoresub{(C,V)}{w}$,
candidate $w$ is not a unique FV winner.

We claim that $\mathcal{S}$ has an exact cover for $B$ if and only if
$w$ can be made a unique FV winner via exerting control by partition
of voters in model~{TP}.

From left to right: Suppose $\mathcal{S}$ contains an exact cover
$\mathcal{S}'$ for $B$. Partition $V$ in the following way. Let $V_1$
consist of:
\begin{itemize}
\item the $m$ voters of the first group that correspond to the exact
  cover (i.e., those $m$ voters of the form
$ \begin{array}{c@{\ \ }c@{\ \ }c@{\ \ }c}
 y & S_i & | & (B-S_i) \cup F \cup Z \cup \{ w,x\}
\end{array}$
for which $S_i \in \mathcal{S}'$),
\item the $n$ voters of the second group (who approve of $y$ and $z_i$ for 
all $i$, $1\leq i \leq n$), and
\item the $n+m+1$ voters of the fifth group (who approve of only $x$).
\end{itemize}
Let $V_2=V-V_1$.

\begin{table}[h]
 \centering
 \begin{tabular}{|c||c|c|c|c|c|c|}
  \hline
                         & $w$ & $x$     & $y$   & $b_{j}$ & $f_k$ & $z_i$ \\ \hline\hline
  $\tabscoresub{(C,V_1)}$  & $0$ & $n+m+1$ & $n+m$ & $1$     & $0$   & $1$   \\ \hline
 \end{tabular}
\caption{Scores in $(C,V_1)$.}	\label{tab:scores-V1}
\end{table}

Table~\ref{tab:scores-V1} shows the overall scores of all candidates
in subelection $(C,V_1)$.
Since the strict majority threshold is $n+m+1$ in subelection $(C,V_1)$, only 
candidate $x$ has a strict majority.  Thus $x$ is the 
unique FV winner of subelection $(C,V_1)$ and is proceeding forward to the 
final round.  

\begin{table}[h]
 \centering
 \begin{tabular}{|c||c|c|c|c|c|c|}
  \hline
                         & $w$  & $x$ & $y$   & $b_{j}$ & $f_k$ & $z_i$    \\ \hline\hline
  $\tabscoresub{(C,V_2)}$  & $2n$ & $0$ & $n-m$ & $2n-1$  & $1$   & $2n-1$   \\ \hline
 \end{tabular}
\caption{Scores in $(C,V_2)$.}	\label{tab:scores-V2}
\end{table}

Table~\ref{tab:scores-V2} shows the overall scores of all candidates
in subelection $(C,V_2)$.  In this subelection, the strict majority
threshold is $2n+1$.  As one can see in Table~\ref{tab:scores-V2},
there is no candidate with a strict majority on any level in election
$(C,V_2)$.  Among all candidates, $w$ has the highest score in
subelection $(C,V_2)$ and thus moves forward to the final round.

Since in the final round none of the two candidates, $x$ and~$w$, has
a strict majority on any level, and since
$\scoresub{(\{w,x\},V)}{w}=2n> n+m+1 =\scoresub{(\{w,x\},V)}{x}$
because $n>m+1$, candidate $w$ is the unique FV winner in the election
resulting from this partition of voters.

From right to left: Suppose that $w$ can be made a unique FV winner by
exerting control by partition of voters in model~{TP}.  Let $V = (V_1,V_2)$
be some such successful partition.
In order to prove that $\mathcal{S}$ has an exact cover for $B$, we will 
show the following three statements:
\begin{enumerate}
\item The final round of the election corresponding to partition $V =
  (V_1,V_2)$ consists of the candidates $x$ and~$w$.
 \item $\mathcal{S}$ has a cover for $B$.
 \item This cover is an exact cover.
\end{enumerate}
 
As to (1), among all candidates, only candidate $x$ and the candidates in $F$
have a score less than $\scoresub{(C,V)}{w}$ (see Table~\ref{tab:scores1}).
Thus, these are the only possible
candidates who could face $w$ in the final 
round.\footnote{Note that in election
$(C,V)$, no candidate has a strict majority. This remains true, if we remove
candidates from $C$. Thus, the winner of the final round is a winner by score
and not by majority on any level.}
Now, a candidate $f_k$ can get into the final round only 
if in $f_k$'s first-round subelection there are only voters
from the sixth group and at most one voter from the fifth
group.\footnote{Otherwise, since 
$\scoresub{(C,V)}{f_k}=1$, it might happen that $y$ or a candidate
$z_i\in Z$ would also be a winner in this subelection and would
move forward to the final round.}
So, if a candidate $f_k$ could make it to the final round then $w$ has
to be the unique FV winner of the other first-round subelection.
However, this is not possible, since both $y$ and $w$ get $2n$ points
in that subelection, and $y$ gains his or her points already on the first
level. Thus, no candidate $f_k\in F$ can participate in the final round.
It follows that the only way to make $w$ a unique FV winner via
exerting control by partition of voters in model~{TP} is that $w$
faces only candidate $x$ in the final round.

As to (2), let $x$ be the unique FV winner of subelection $(C,V_1)$ and 
$w$ the unique FV winner of subelection $(C,V_2)$.
Since in each vote in the third and fourth voter group, each candidate
$b_j\in B$ appears on an earlier level than~$w$, there has to be a cover
in $\mathcal{S}$ for~$B$, for otherwise there would be at least one 
$b_j\in B$ who ties $w$ in $(C,V_2)$ by score, and reaches that score on 
an earlier level than~$w$.  Let the size of the cover be~$m'$.
Note that $m'\geq m$.

As to (3), note that each voter of the second group has to be in
$(C,V_1)$, for otherwise there would be at least one $z_i\in Z$ who
ties $w$ in $(C,V_2)$ by score, and reaches that score on an earlier
level than~$w$.  The following must hold:
\[
 \scoresub{(C,V_1)}{x}-\scoresub{(C,V_1)}{y}=(n+m+1)-(n+m')=m+1-m'>0.
\]
This is possible only if $m'=m$. Hence, $\mathcal{S}$ has an
\emph{exact} cover for~$B$, which completes the proof.~\end{proofs}

The following construction will be used to handle the destructive case
of control by partition of voters in model~{TP}.

\begin{construction}
\label{con:destructive-partition-voters-TP}
Let $(B,\mathcal{S},k)$ be a given instance of $\hittingset$, where $B
= \{b_1, b_2, \ldots , b_m\}$ is a set, $\mathcal{S} = \{S_1, S_2,
\ldots , S_n\}$ is a collection of nonempty subsets $S_i \seq B$, and
$k<m$ is a positive integer.\footnote{Note that the assumption $k<m$
can be made without loss of generality, since the problem $\hittingset$
becomes trivial if $k=m$.}

Define the election $(C,V)$, where $C = B \cup D \cup E \cup \{c,w\}$ is the
candidate set with $D=\{ d_1, \ldots , d_{2(m+1)}\}$ and
$E=\{e_1, \ldots , e_{2(m-1)}\}$ 
and where $V$ consists  of the following $2n(k+1)+4m+2mk$ voters:
\begin{enumerate}
\item For each $i$, $1\leq i \leq n$, there are $k+1$ voters of the
  form:
\[
\begin{array}{c@{\ \ }c@{\ \ }c@{\ \ }c@{\ \ }c}
w & S_i & c & | & D\cup E \cup (B-S_i).
\end{array}
\]
\item For each $j$, $1\leq j \leq m$, there is one voter of the form:
\[
\begin{array}{c@{\ \ }c@{\ \ }c@{\ \ }c@{\ \ }c}
c & b_j & w & | & (B-\{b_j\})\cup D \cup E.
\end{array}
\]
\item For each $j$, $1\leq j \leq m$, there are $(k-1)$ voters of the form:
\[
\begin{array}{c@{\ \ }c@{\ \ }c}
b_j & | & (B-\{b_j\})\cup D \cup E \cup \{c,w \}.
\end{array}
\]
\item For each $p$, $1 \leq p \leq m+1$, there is one voter of the form:
\[
\begin{array}{c@{\ \ }c@{\ \ }c@{\ \ }c@{\ \ }c}
 d_{2(p-1)+1} & d_{2p} & w & | & B\cup (D-\{d_{2(p-1)+1},d_{2p}\}) \cup E \cup \{c \}.
\end{array}
\]
\item For each $r$, $1 \leq r \leq 2(m-1)$, there is one voter of the form:
\[
\begin{array}{c@{\ \ }c@{\ \ }c}
e_r & | & B\cup D \cup (E-\{ e_r \} \cup \{c,w \}.
\end{array}
\]
\item There are $n(k+1)+m-k+1$ voters of the form:
\[
\begin{array}{c@{\ \ }c@{\ \ }c}
 c & | & B\cup D\cup E \cup \{w\}.
\end{array}
\]
\item There are $mk+k-1$ voters of the form:
\[
\begin{array}{c@{\ \ }c@{\ \ }c@{\ \ }c}
 c & w & | & B\cup D \cup E.
\end{array}
\]
\item There is one voter of the form:
\[
\begin{array}{c@{\ \ }c@{\ \ }c@{\ \ }c}
 w & c & | & B\cup D \cup E.
\end{array}
\]
\end{enumerate}
\end{construction}

The strict majority threshold for $V$ is $\mbox{maj}(V) = n(k+1)+2m +
mk +1$.  In election $(C,V)$, only the two candidates $c$ and $w$ reach
a strict majority, $w$ on the third level and $c$ on the second level
(see Table~\ref{scores(C,V)}).  Thus $c$ is the unique level $2$ FV
winner of election $(C,V)$.

\begin{table}
\centering
{\small
\begin{tabular}{|l||c|c|c|c|c|}
\hline
                   & $c$                   & $w$                & $b_{j}$         & $d_p$    & $e_r$ \\\hline\hline
$\mbox{score}^{1}$ & $ n(k+1)+2m +mk$      & $n(k+1)+1$         & $k-1$           & $\leq 1$ & $1$   \\\hline
$\mbox{score}^{2}$ & $ n(k+1)+2m+mk+1$     & $n(k+1)+mk+k$      & $\leq k+n(k+1)$ & $1$      & $1$   \\\hline
$\mbox{score}^{3}$ & $\geq n(k+1)+2m+mk+1$ & $n(k+1)+mk+k+2m+1$ & $\leq k+n(k+1)$ & $1$      & $1$   \\\hline
\end{tabular}
}
\caption{Scores in $(C,V)$.}
\label{scores(C,V)}
\end{table}

The proof of Theorem \ref{thm:destructive-partition-voters-TP} will
make use of the following claim.

\begin{claim}
\label{cla:destructive-partition-voters-TP}
In election $(C,V)$ from
Construction~\ref{con:destructive-partition-voters-TP}, for every
partition of $V$ into $V_1$ and $V_2$, candidate $c$ is an FV winner
of either $(C,V_1)$ or $(C,V_2)$.
\end{claim}

\begin{proofs}
For a contradiction, suppose that in both subelections, 
$(C,V_1)$ and $(C,V_2)$, candidate $c$ is not an FV winner. 
In particular, $c$ can have no strict majority in either of 
$(C,V_1)$ and $(C,V_2)$. 
Since $\scoresublevel{(C,V)}{1}{c}= \nicefrac{\|V\|}{2}$, the two subelections 
must satisfy the following conditions:
\begin{enumerate}
	\item Both $\|V_{1}\|$ and $\|V_{2}\|$ are even numbers and
	\item $\scoresublevel{(C,V_1)}{1}{c}= \nicefrac{\|V_1\|}{2}$ and
		$\scoresublevel{(C,V_2)}{1}{c}= \nicefrac{\|V_2\|}{2}$.
\end{enumerate}

Otherwise, $c$ would have a strict majority already on the first level
in one of the subelections. Since in both 
subelections $c$ has only one point less than the 
strict majority threshold already on the first level, and since 
$c$ will get a strict majority no 
later than on the second level, in both subelections there must be 
candidates whose level $2$ scores are higher than the level 
$2$ score of candidate~$c$. 
Table~\ref{scores(C,V)} shows the level $2$ scores of all 
candidates. Only candidates $w$ and a $b_j\in B$ have a chance to 
tie or beat candidate $c$ on that level.

Essentially, there are two possibilities for winning the two subelections. 
First, it is possible that 
both subelections are won by two distinct candidates from $B$ 
(say, $b_x$ is a winner of $(C,V_1)$ and $b_y$ is a winner of $(C,V_2)$). 
Thus the following must hold:

\begin{eqnarray*}
\scoresublevel{(C,V_1)}{2}{b_x} + \scoresublevel{(C,V_2)}{2}{b_y} &
 \geq & \scoresublevel{(C,V)}{2}{c} \\
                                                2n(k+1)+2k-n(k+1) &
 \geq & n(k+1)+mk+2m+1              \\
                                                               2k &
 \geq & mk+2m+1                     \\
                                                                0 &
 \geq & (m-2)k+2m+1.
\end{eqnarray*}
This is a contradiction to the basic assumption that both 
$k>0$ and $m>0$.  Thus only the second possibility for $c$ to lose 
both subelections remains, namely that one subelection, say $(C,V_1)$,
is won by a candidate from~$B$, say~$b_x$, and the other subelection,
$(C,V_2)$, is won by candidate~$w$.  Then it must hold that:
\begin{eqnarray*}
\scoresublevel{(C,V_1)}{2}{b_x} + \scoresublevel{(C,V_2)}{2}{w} &
 \geq & \scoresublevel{(C,V)}{2}{c} \\
                                    n(k+1)+k-n(k+1) + n(k+1)+mk+k &
 \geq & n(k+1)+mk+2m+1              \\
                                                             2k &
 \geq & 2m+1.
\end{eqnarray*}

This is a contradiction to the assumption that $k<m$, so $c$ 
must be an FV winner in one of the subelections.~\end{proofs}

\begin{theorem}
 \label{thm:destructive-partition-voters-TP}
Fallback voting is resistant to destructive control by 
partition of voters in model~TP. 
\end{theorem}

\begin{proofs}
Susceptibility holds by Lemma~\ref{lem:susceptible-voter-control}. 
To prove $\np$-hardness, we reduce Hitting Set to our control problem. 
Consider the election $(C,V)$ constructed according to 
Construction~\ref{con:destructive-partition-voters-TP} from a given
Hitting Set instance $(B,\mathcal{S},k)$, where 
$B=\{b_1,\ldots,b_m\}$ is a set, $\mathcal{S}=\{S_1,\ldots,S_n\}$ is 
a collection of nonempty subsets $S_i\subseteq B$, and $k<m$ is a positive 
integer. 

We claim that $\mathcal{S}$ has a hitting set $B'\subseteq B$ of 
size $k$ if and only if $c$ can be prevented from winning by partition 
of voters in model~TP.

From left to right: Suppose, $B' \subseteq B$ is a hitting 
set of size $k$ for $\mathcal{S}$. Partition $V$ into $V_1$ and $V_2$
the following way. 
Let $V_1$ consist of those voters of the second group where $b_j\in B'$ 
and of those voters of the third group where $b_j\in B'$.
Let $V_2 = V-V_1$. In $(C,V_1)$, no candidate reaches a strict majority
(where $\mbox{maj}(V_1) = \|B'\| + 1 = k+1$),
and candidates $c$, $w$, and $b_j\in B'$ win the election with an approval 
score of $k$ (see Table~\ref{scores(E,V1)}). 

\begin{table}[h!t]
\centering
\begin{tabular}{|l||c|c|c|c|}
\hline
                            & $c$ & $w$ & $b_{j}\in B'$ & $b_{j}\not\in B'$ \\\hline\hline
	 $\mbox{score}^{1}$ & $k$ & $0$ & $k-1$         & $0$               \\\hline
	 $\mbox{score}^{2}$ & $k$ & $0$ & $k$           & $0$               \\\hline
	 $\mbox{score}^{3}$ & $k$ & $k$ & $k$           & $0$		    \\
\hline	 
\end{tabular}
\caption{Scores in $(C,V_1)$.}
\label{scores(E,V1)}
\end{table}

The scores in election $(C,V_2)$ are shown in Table~\ref{scores(E,V2)}. 

\begin{table}[h!t]
\centering
{\small
	\begin{tabular}{|l||c|c|c|c|}
\hline
	                    & $c$                     & $w$                & $b_{j}\not\in B'$ & $b_{j}\in B'$ \\\hline\hline
	 $\mbox{score}^{1}$ & $n(k+1)+2m-k+mk$        & $n(k+1)+1$         & $k-1$             & $0$           \\\hline
	 $\mbox{score}^{2}$ & $n(k+1)+2m-k+mk+1$      & $n(k+1)+mk+k$      & $\leq k+n(k+1)$   & $\leq n(k+1)$ \\\hline
	 $\mbox{score}^{3}$ & $\geq n(k+1)+2m-k+mk+1$ & $n(k+1)+mk+2m+1$   & $\leq k+n(k+1)$   & $\leq n(k+1)$ \\
\hline
	\end{tabular}
}
\caption{Scores in $(C,V_2)$.}
\label{scores(E,V2)}
\end{table}

Since in election $(C,V_2)$ no candidate from $B$ wins, the 
candidates participating in the final round are $B'\cup\{c,w\}$. 
The scores in the final election $(B'\cup\{c,w\},V)$ can be seen in 
Table~\ref{scoresFinal}.
Since candidates $c$ and $w$ are both level $2$ FV winners,
candidate $c$ is no longer the unique FV winner of the election.

\begin{table}[h!t]
\centering
{\small
	\begin{tabular}{|l||c|c|c|}
\hline
	                    & $c$              & $w$               & $b_{j}\in B'$   \\\hline\hline
	 $\mbox{score}^{1}$ & $n(k+1)+2m+mk$   & $n(k+1)+m+2$   & $k-1$              \\\hline
	 $\mbox{score}^{2}$ & $n(k+1)+2m+mk+1$ & $n(k+1)+2m+mk+1$  & $\leq k+n(k+1)$ \\ 
\hline
	\end{tabular}
}
\caption{Scores in $(B'\cup\{c,w\},V)$.}
\label{scoresFinal}
\end{table}

From right to left: Suppose candidate $c$ can be prevented 
from winnning by partition of voters in model~TP. From 
Claim~\ref{cla:destructive-partition-voters-TP} it follows that 
candidate $c$ participates in the final round. For a contradiction, 
suppose that $\mathcal{S}$ has not a hitting set of size $k$. Since 
$c$ has a strict majority of approvals, $c$ has to be tied with or 
lose against another candidate by a strict majority at some level. 
Only candidate $w$ has a strict majority of approvals, so $w$ has to 
tie or beat $c$ at some level in the final round. 
Because of the scores of the candidates from $D$ and $E$ we may assume 
that only candidates from $B$ are participating in the final round 
besides $c$ and $w$. Let $B'\subseteq B$ be the set of candidates who 
also participate in the final round and let $\ell$ be the number of sets 
in $\mathcal{S}$ not hit by $B'$. Note that $w$ cannot reach a strict 
majority of approvals on the first level, so we consider the level 
$2$ scores of $c$ and $w$: 

\begin{eqnarray*}
\scoresublevel{(B'\cup\{c,w\},V)}{2}{c} & = & n(k+1)+2m+mk+1+\ell(k+1)
\quad \text{ and} \\
\scoresublevel{(B'\cup\{c,w\},V)}{2}{w} & = & n(k+1)+2m+mk+k-\|B'\|+1.
\end{eqnarray*}

Since $c$ has a strict majority already on the second level, 
$w$ must tie or beat $c$ on this level, so the following must hold:

\begin{eqnarray*}
\scoresublevel{(B'\cup\{c,w\},V)}{2}{c} - 
\scoresublevel{(B'\cup\{c,w\},V)}{2}{w} & \leq & 0 \\
n(k+1)+2m+mk+1+\ell(k+1)- n(k+1)-2m-mk-k+\|B'\|-1 & \leq & 0 \\
\|B'\|-k+\ell(k+1) & \leq & 0 .
\end{eqnarray*}

This is possible only if $\ell=0$, which contradicts to our assumption
that there are sets in $\mathcal{S}$ that are not hit by~$B'$.  From
$\ell=0$ it follows that $\|B'\| \leq k$, so $\mathcal{S}$ has a
hitting set of size at most~$k$.~\end{proofs}

\section{Conclusions and Open Questions}
\label{sec:conclusions}

We have shown that Brams and Sanver's fallback voting
system~\cite{bra-san:j:preference-approval-voting} is, like plurality
voting and SP-AV, fully resistant to candidate control.  Also, like
Copeland
voting~\cite{fal-hem-hem-rot:j:llull-copeland-full-techreport} and
SP-AV, fallback voting is fully resistant to constructive control.
Regarding voter control, all eight cases in FV are susceptible, and we
have shown resistance to constructive control by adding, by deleting,
and by partition of voters in models TE and TP, and by destructive
control to partition of voters in model~{TP}.  We have also shown
vulnerability to destructive control by adding and by deleting voters.
Only one case
remains open: destructive control by partition of
voters in model~{TE}.  It would be interesting to know whether FV
is resistant or vulnerable to this control type.

Plurality voting is one of the other two systems for which full
resistance to candidate control is known
\cite{hem-hem-rot:j:destructive-control}, but it has fewer resistances
to voter control than fallback voting.  SP-AV (the other system with
known full resistance to candidate control) does have
the same number
of proven resistances \cite{erd-now-rot:j:sp-av} to voter control
as fallback voting.  However, as has been argued in the introduction, it
is less natural a system than fallback voting.  Also, it is still
possible that fallback voting might turn out to have even one more
resistance to control than SP-AV in total.

Of course, resistance to control is not the only---and probably not
even the most important---criterion to guide one's choice of voting
system.  Many other properties of voting systems (especially their
social choice weaknesses and strengths) are important as well and
perhaps even more important.  For example, representing votes in
plurality is a slightly simpler task than in fallback voting or SP-AV:
Plurality voters simply give a ranking of the candidates and the
candidates with the most top positions win, whereas fallback
and SP-AV voters provide both their approvals/disapprovals of the
candidates and a ranking of the candidates (of all candidates in SP-AV
and of only the approved candidates in fallback voting).  Also, winner
determination in fallback voting and in SP-AV is a slightly more
complicated task than in plurality voting---though still easy.
Regarding the social choice benefits of FV, we mention that it
satisfies, e.g., monotonicity and refer to
\cite{bra-san:j:preference-approval-voting} for a more detailed
discussion and further interesting results.

Supposing one does care about control resistance, when choosing a
voting system one's choice will most likely (along with the system's
social choice properties, of course) depend on the types of control
one cares most about in the intended application.
Also, when comparing voting systems,
one should weigh the nine immunities, four resistances, and nine
vulnerabilities to control approval voting is known to possess
\cite{hem-hem-rot:j:destructive-control} against FV's at least
$19$ (and possibly even~$20$) resistances and at least two (and
at most three) vulnerabilities to control.

{\small 
\bibliographystyle{alpha}

\bibliography{fv}
}

\appendix

\section{Some Results of \cite{hem-hem-rot:j:destructive-control} 
Used in Section~\ref{sec:results:susceptibility}}

\begin{theorem}[\cite{hem-hem-rot:j:destructive-control}]
 \label{thm:voiced-control}
 \begin{enumerate}
 \item If a voiced voting system is susceptible to destructive control
	by partition of voters (in model TE or TP), 
	then it is susceptible to destructive control by deleting voters.
 \item Each voiced voting system is susceptible to constructive control
	by deleting candidates.
 \item Each voiced voting system is susceptible to destructive control
	by adding candidates.\footnote{Following
Bartholdi et al.~\cite{bar-tov-tri:j:control},
Hemaspaandra et al.~\cite{hem-hem-rot:j:destructive-control}
considered only the case of control by adding a limited number of
candidates---the ``unlimited'' case was introduced only in (the conference
precursors of) \cite{fal-hem-hem-rot:j:llull-copeland-full-techreport}.
However, it is easy to see that all results about control by adding
candidates stated in
Theorems~\ref{thm:voiced-control},
\ref{thm:duality-constructive-destructive-control},
and~\ref{thm:susceptibility-implications} hold true in both the limited
and the unlimited case.}
\end{enumerate}
\end{theorem}

\begin{theorem}[\cite{hem-hem-rot:j:destructive-control}]
 \label{thm:duality-constructive-destructive-control}
 \begin{enumerate}
 \item A voting system is susceptible to constructive control by adding
candidates if and only if it is susceptible to destructive control by
deleting candidates.
 \item A voting system is susceptible to constructive control by deleting
candidates if and only if it is susceptible to destructive control by
adding candidates.
 \item A voting system is susceptible to constructive control by adding
voters if and only if it is susceptible to destructive control by
deleting voters.
 \item A voting system is susceptible to constructive control by deleting
voters if and only if it is susceptible to destructive control by
adding voters.
\end{enumerate}
\end{theorem}

\begin{theorem}[\cite{hem-hem-rot:j:destructive-control}]
 \label{thm:susceptibility-implications}
\begin{enumerate}
\item If a voting system is susceptible to constructive control by
partition of voters (in model TE or TP), then it is susceptible to
constructive control by deleting candidates.

\item If a voting system is susceptible to constructive control by
partition or run-off partition of candidates (in model TE or TP), then
it is susceptible to constructive control by deleting candidates.

\item If a voting system is susceptible to constructive control by
partition of voters in model TE, then it is susceptible to
constructive control by deleting voters.

\item If a voting system is susceptible to destructive control by
partition or run-off partition of candidates (in model TE or TP), then
it is susceptible to destructive control by deleting candidates.
\end{enumerate}
\end{theorem}

\end{document}